# Macromammal and bird assemblages across the late Middle to Upper Palaeolithic transition in Italy: an extended zooarchaeological review

Matteo Romandini[a,b,1,*], Jacopo Crezzini[c,1], Eugenio Bortolini[a,1], Paolo Boscato[c], Francesco Boschin[c], Lisa Carrera[d], Nicola Nannini[e], AntonioTagliacozzo[f], GabrieleTerlato[b], Simona Arrighi[a,c], Federica Badino[a,g], Carla Figus[a], Federico Lugli[a], Giulia Marciani[a], Gregorio Oxilia[a], Adriana Moroni[c], Fabio Negrino[h], Marco Peresani [b], JulienRiel-Salvatore[i], Annamaria Ronchitelli[c], Enza Elena Spinapolice[j], Stefano Benazzi[a]

[a] *Università di Bologna, Dipartimento di Beni Culturali, Via degli Ariani 1, 48121, Ravenna, Italy*
[b] *Università di Ferrara, Dipartimento di Studi Umanistici, Sezione di Scienze Preistoriche e Antropologiche, Corso Ercole I d'Este 32, 44100, Ferrara, Italy*
[c] *Università di Siena, Dipartimento di Scienze Fisiche, della Terra e dell'Ambiente, U.R. Preistoria e Antropologia, Via Laterina 8, 53100, Siena, Italy*
[d] *Università di Bologna, Dipartimento di Scienze Biologiche, Geologiche e Ambientali, Via Zamboni 67, 40126, Bologna, Italy*
[e] *MuSe - Museo delle Scienze, Corso del Lavoro e della Scienza 3, I-38123, Trento, Italy*
[f] *Service of Bioarchaeology, Museo delle Civiltà, Museo Preistorico Etnografico "Luigi Pigorini", Piazzale G. Marconi 14, I-00144, Rome, Italy*
[g] *C.N.R, Istituto di Geologia Ambientale e Geoingegneria, 20126, Milano, Italy*
[h] *Università di Genova, Dipartimento di Antichità, Filosofia, Storia, Via Balbi 2, 16126, Genova, Italy*
[i] *Université de Montréal, Département d'Anthropologie, 2900, Edouard Montpetit Blvd, Montréal, QC, H3T 1J4, Canada*
[j] *Sapienza University of Rome, Dipartimento di Scienze dell'Antichità, Piazzale Aldo Moro 5, 00185, Roma, Italy*

A R T I C L E I N F O

A B S T R A C T

Evidence of human activities during the Middle to Upper Palaeolithic transition is well represented from Protoaurignacian rock-shelters, caves and open-air sites across Italy.
Over the past decade, both the revision of taphonomic Uluzzian processes affecting archaeological faunal assemblages and new zooarchaeological studies have allowed archaeolLate mousterian ogists to better understand subsistence strategies and cultural behaviors attributed to groups of Neandertal and Zooarchaeology modern humans living in the region. This work presents the preliminary results of a 5-years research programme Aoristic analysis (ERC n. 724046 – SUCCESS) and offers a state-of-the-art synthesis of archaeological faunal assemblages including mammals and birds uncovered in Italy between 50 and 35 ky ago. The present data were recovered in primary Late Mousterian, Uluzzian, and Protoaurignacian stratigraphic contexts from Northern Italy (Grotta di Fumane, Riparo del Broion, Grotta Maggiore di San Bernardino, Grotta del Rio Secco, Riparo Bombrini), and Southern Italy (Grotta di Castelcivita, Grotta della Cala, Grotta del Cavallo, and Riparo l'Oscurusciuto). The available Number of Identified Specimens (NISP) is analysed through intra- and inter-site comparisons at a regional scale, while aoristic analysis is applied to the sequence documented at Grotta di Fumane. Results of qualitative comparisons suggest an increase in the number of hunted taxa since the end of the Middle Palaeolithic, and a marked change in ecological settings beginning with the Protoaurignacian, with a shift to lower temperatures and humidity. The distribution of carnivore remains and taphonomic analyses hint at a possible change in faunal exploitation and butchering processing between the Middle and Upper Palaeolithic. A preliminary comparison between bone frequencies and the distribution of burned bones poses interesting questions concerning the management of fire. Eventually, the combined use of relative taxonomic abundance and aoristic analysis explicitly addresses time averaging and temporal uncertainty embedded in NISP counts and offers estimates of absolute change over time that can be used to support hypotheses emerging from taxon relative frequencies.

* Corresponding author. Università di Bologna, Dipartimento di Beni Culturali, Via degli Ariani 1, 48121, Ravenna, Italy.

*E-mail addresses:* matteo.romandini@unibo.it (M. Romandini); jacopocrezzini@gmail.com (J. Crezzini); eugenio.bortolini2@unibo.it (E. Bortolini); paolo.boscato@unisi.it (P. Boscato); fboschin@hotmail.com (F. Boschin); lisa.carrera3@unibo.it (L. Carrera); nicola.nannini@muse.it (N. Nannini); antonio.tagliacozzo@beniculturali.it (A. Tagliacozzo); gabriele.terlato@ unife.it (G. Terlato); simona.arrighi@unibo.it (S. Arrighi); federica.badino@unibo.it (F. Badino); carla.figus3@unibo.it (C. Figus); federico.lugli6@unibo.it (F. Lugli); giulia.marciani@ unibo.it (G. Marciani); gregorio.oxilia3@unibo.it (G. Oxilia); adriana.moroni@unisi.it (A. Moroni); fabio.negrino@unige.it (F. Negrino); marco.peresani@unife.it (M.); julien.rielsalvatore@umontreal.ca (J. Riel-Salvatore); annamaria.ronchitelli@unisi.it (A. Ronchitelli); enzaelena.spinapolice@uniroma1.it (E.E. Spinapolice); stefano.benazzi@unibo.it (S. Benazzi) [1] these authors equally contributed to the present work.

## 1. Introduction

Evidence for change in human behaviour and adaptive strategies linked to palaeoenvironmental change has been consistently documented for contexts dated to Marine Isotope Stage 3 (MIS 3: 60-30 ky BP) across Europe. The different subsistence strategies developed by Neandertals and modern humans in response to change in the underlying climatic conditions has been of particular interest in all transitional contexts of continental and Mediterranean Europe (among others: Bietti and Manzi, 1990-91; Guidi and Piperno, 1992; Stiner, 1994; Bietti and Grimaldi, 1996; Milliken, 1999–2000; Kuhn and Bietti, 2000; Mussi, 2001; Peresani, 2009, 2011; Moroni et al., 2013, 2019).

Investigations into hominin diets, specifically those of the Neandertals, ineluctably feed into debates that revolve around the presumed capabilities, or lack thereof, of these hominins in the exploitation of small game as a food resource (Stiner, 2001; Stiner and Munro, 2002, 2011; Hockett and Haws, 2005).

Nevertheless, multiple data have induced some authors to suggest that the exploitation of small animals has been important for human subsistence since ca. 250ka (Klein and Scott, 1986; Stiner, 2005; Romandini et al., 2018b; Morin et al., 2019).

The Italian Peninsula plays a pivotal role as it connects Alpine Europe to the centre of the Mediterranean, and it provides a privileged perspective on interaction and replacement of Neandertals by modern humans in a very diverse set of ecological and climatic regions (Benazzi et al., 2011; Higham et al., 2011; Peresani, 2011; Moroni et al., 2018; Villa et al., 2018; Peresani et al., 2016, 2019). All scholars agree for example on the role played by geographic barriers (Alps and Apennines) in segregating – from a climatic and ecological point of view – a western Mediterranean region form an eastern continental one, the latter affected by the cyclical emersion of the northern Adriatic platform (Sala, 1990; Sala and Marchetti, 2006). Such a diversity, however, made the reconstruction of past ecosystems, of the spatio-temporal distribution of resources, and of population-level subsistence strategies particularly difficult, especially in light of the intense glacial/interglacial cycles of the past 200,000 years. Notwithstanding the many detailed studies carried out at a local scale, a global understanding of change in mobility, adaptive strategies, and settlement pattern across the Middle-Upper Palaeolithic Transition across Italy is still elusive. The few exceptions (Van Andel and Davies, 2003) draw on very scant and heterogeneous data generated with different aims and at different scales, and the emerging scenarios are far from the temporal coherence exhibited by recent global (Bond et al., 1992; Dansgaard et al., 1993; Rasmussen et al., 2014) and Mediterranean palaeoclimatic and palaeoecological records (Allen et al., 1999; Sánchez Goñi et al., 2000; Tzedakis et al., 2002; Margari et al., 2009; Fletcher et al., 2010; Müller et al., 2011; Wulf et al., 2018).

The present paper aims to fill this gap and lay the foundations for a finer and more systematic comparison across the whole of the Italian Peninsula by presenting a state-of-the-art review of available data on faunal remains in a number of key dated Italian sites. By carefully documenting and comparing the distribution of faunal remains, we also generate hypotheses on the different subsistence strategies developed by Neandertals and modern humans in response to change in the underlying climatic conditions. Inferences about paleoclimate and ecological settings are based on well-established links between ungulate families/avifaunal groups and the very specific environmental settings to which they were and still are adapted today. More specifically, the review focuses on relative taxon frequency of macromam mals (ungulates, carnivores, rodents and lagomorphs) and birds across Late Mousterian, Uluzzian, and Protoaurignacian layers documented for 9 Italian sites for which quantitative data are available (Fig. 1, Areas 1–3). Other assemblages from central and northwestern Italy are also briefly described in this context, but their data are not directly integrated in more detailed investigations of regional trends. Finally, one particularly well-documented site (Grotta di Fumane) is also investigated through aoristic analysis, a probabilistic approach never before applied to the Palaeolithic of Italy. The method explicitly addresses temporal uncertainty and depositional factors affecting the observed number of specimens (NISP) and offers estimates of absolute change over time that can be used to support hypotheses emerging from taxon relative frequencies, as well as to allow a direct comparison between layers of different coeval sites. Finally, the available taphonomic evidence is also presented to provide preliminary insights on change over time in animal exploitation strategies and butchering processes.

This work is still preliminary, as it describes the initial results of an ongoing 5-year project aimed at reaching a deeper understanding of the mechanisms that underpinned the geographic overlap between Neandertals and modern humans in the Italian Peninsula, as well as the final replacement of the former by the latter. While future research will be able to support or disprove part of the picture that emerges from this first assessment, it nonetheless offers a first attempt to generate a coherent synthesis of all the data published to date concerning the region of interest.

## 2. Regional contexts

### *2.1. Northeastern (Adriatic) Italy*

In the northern Adriatic Area archaeologists uncovered a considerable number of rock shelters and caves which yielded evidence of the last Neandertals and of the earliest modern humans. The geographic location of such sites is a key element to understanding regional differences in the faunal assemblages they have yielded (Sala, 1990; Sala and Marchetti, 2006; Sala and Masini, 2007; Masini and Sala, 2007, 2011). From a paleoecological point of view, pollen records from Lake Fimon and Azzano Decimo (north-eastern Alpine foothills; Pini et al., 2009, 2010), document long-term vegetation trends during MIS 3. Phases of expansion of conifer-dominated forest (*Pinus sylvestris/mugo* and *Picea*), rich in broad-leaved trees (*Alnus* cf. *incana* and tree *Betula*), are accompanied by a reduction in the amount of warm-temperate elements (e.g. *Tilia*). Middle Würm stadials experienced summer temperatures very close to the growth limit of oaks, but still within the range of lime (MAW 13–15 °C) which persisted, together with other temperate trees (e.g. *Abies*), up to ca. 40 ka (Pini et al., 2009, 2010; Badino et al., this Special Issue). Interestingly, peaks of *Tilia* pollen have been identified in layers preserving Mousterian artifacts and dated to 40.6–46.4 ka $^{14}$C BP from cave deposits at the Broion (Leonardi and Broglio, 1966; Cattani and Renault-Miskowski, 1983-84). Despite evidence of afforestation persisting at a long-term scale south of the Alps, forest withdrawals with expansion of grasslands and dry shrublands (Gramineae, *Artemisia*, Chenopodiaceae) occurred, possibly related to the establishment of drier/colder conditions (i.e. Greenland Stadials/Heinrich events). Such drier and colder stadial conditions likely favoured the presence of Alpine ibex, chamois, and marmot at low altitudes (in the Colli Berici), as well as the presence of micromammals in steppic environments, and the diffusion of birds in tundra-like environments. The Po alluvial valley was, in contrast, inhabited by woolly rhinoceros, mammoth, and bison (Sala, 1990).

Only a few contexts offer data on the Middle to Upper Palaeolithic transition, and their number further decreases for the temporal interval comprised between 50 and 35 ky.

At present, data on faunal remains and the relative chronology are available from Grotta di Fumane (Verona), Grotta Maggiore di San Bernardino (Vicenza), Riparo del Broion and Grotta del Broion (Vicenza), Grotta del Rio Secco (Pordenone) (Table 1, Fig. 1, Area 1).

**Grotta di Fumane** is a key site for northern Italy, located at 350 m asl in the western part of the Lessini Mountains (Table 1, Fig. 1). Its archaeological sequence includes the Middle-Upper Palaeolithic transition (Peresani et al., 2008; Higham et al., 2009; López-García et al., 2015). Faunal assemblages consist of a rich association of ungulates, carnivores, and birds from diverse environments and climates. Quantitative comparisons between the Uluzzian assemblage (A3) (Tagliacozzo et al., 2013) and the late Mousterian ones (A4, A5-A6, A9) has highlighted only modest ecological and economic adjustments within a humid forested landscape (Peresani et al., 2011a,b; Romandini, 2012; Romandini et al., 2014a, 2016a,b, 2018a,b, Fiore et al., 2016; Gala et al., 2018; Terlato et al., 2019). Considerable change, on the other hand, emerged from the Protoaurignacian occupations (A2), coinciding with a shift towards colder and steppic environments (Cassoli and Tagliacozzo, 1994a; Fiore et al., 2004).

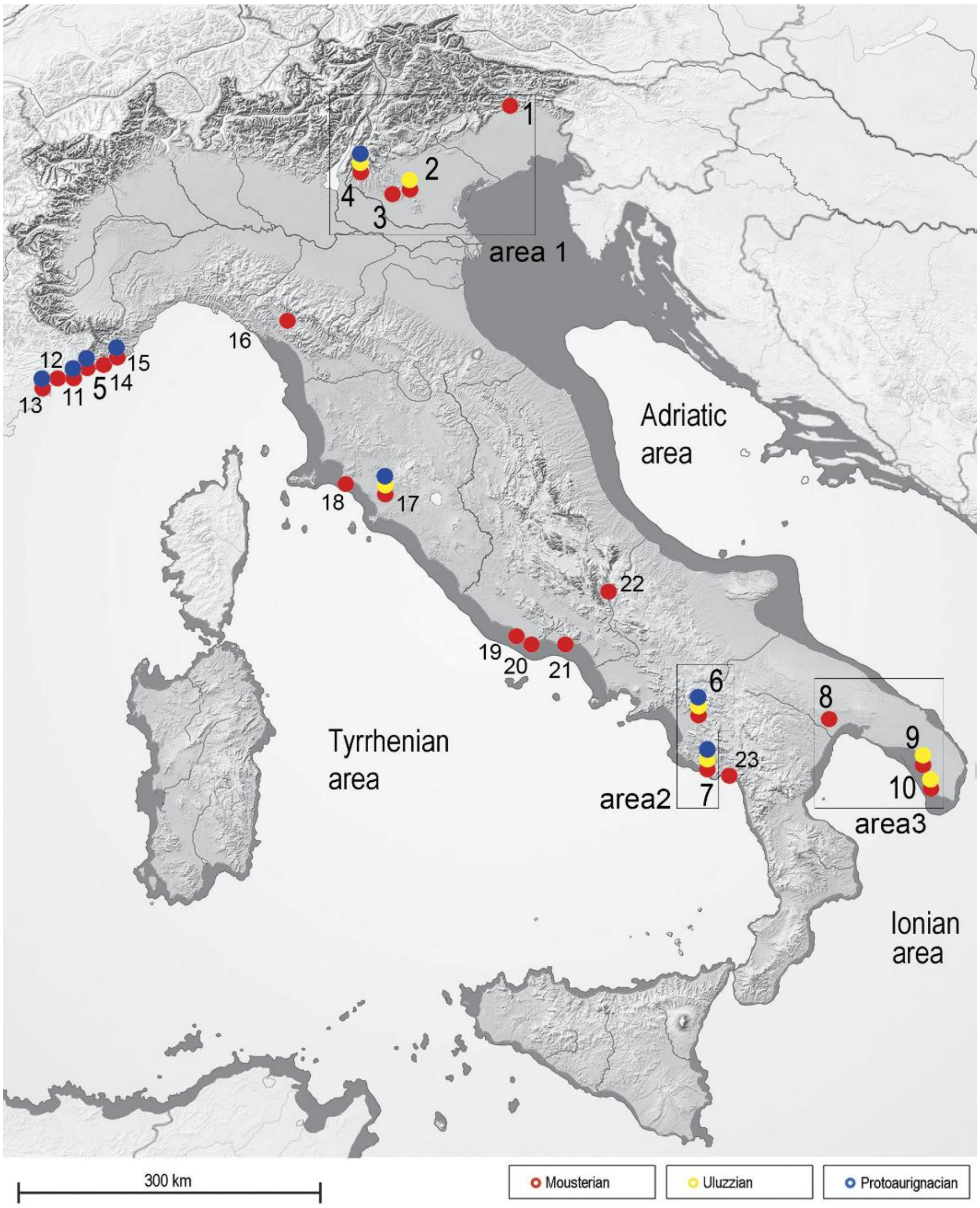

**Fig. 1.** MIS 3 map of Italy (modified from Moroni et al., 2018) and the geographic location of the sites with previously published faunal assemblages mentioned in this work and dated between ca. 50 and 38 ky ago. Sites that are part of the project ERC n. 724046 ‐ SUCCESS are numbered from 1 to 10. Sites analysed in this work are assigned numbers 1‐9 (with the exclusion of 5), and are located in study Areas 1‐3 (Northeastern, Southwester/Tyrrhenian, and Southeastern/Ionian respectively). For each site the colors represent the presence of levels, USS and/or layers chronologically and technologically linked respectively to the Protoaurignacian = blue; Uluzzian = yellow; Late Mousterian = red. 1) Grotta del Rio Secco; 2) Riparo del Broion; 3) Grotta di San Bernardino; 4) Grotta di Fumane; 5) Riparo Bombrini; 6) Grotta di Castelcivita; 7) Grotta della Cala; 8) Riparo l'Oscurusciuto; 9) Grotta del Cavallo; 10) Grotta di Uluzzo C; 11) Riparo Mochi; 12) Grotta del Principe; 13) Observatoire; 14) Arma Manie; 15) Arma degli Zerbi; 16) Buca della Iena; 17) Grotta la Fabbrica; 18) Grotta dei Santi; 19) Grotta Breuil; 20) Grotta del Fossellone; 21) Grotta S. Agostino; 22) Grotta Reali; 23) Riparo del Poggio. The Italian Peninsula shows a sea level of 70 m below the present-day coastline, based on the global sea-level curve (Benjamin et al., 2017) but lacking the estimation of post-MIS3 sedimentary thickness and eustatic magnitude (sketch map courtesy of S. Ricci, University of Siena). (For interpretation of color coding in this figure legend, the reader is referred to the Web version of this article.)

| Region | Sites | US/levels | Technocomplex | C14 cal BP | U/Th | Tot. NISP Ungulates | Dominant taxa | Climate/Enviroment |
|---|---|---|---|---|---|---|---|---|
| Northern Italy | RS-Rio Secco | 5+8 | Late Mousterian | >48-44 ky BP | - | 42 | *Ursus* sp. | cold-temperate climate with humid condition and open enviroments |
| | RF-Fumane | A9 | Late Mousterian | 47-45 ky BP | | 1214 | | |
| | RF-Fumane | A6 | Late Mousterian | 44-42 ky BP | - | 1570 | *Cervus elaphuhs* + *Capreolus capreolus* | temperate climate with forests and clearings |
| | RF-Fumane | A5/A5+A6 | Late Mousterian | | - | 479 | | |
| | RS-Rio Secco | 5top+7 | Late Mousterian | >49-41 ky BP | - | 58 | *Ursus* sp. | cold-temperate climate with humid condition and open enviroments |
| | SB-S. Bernardino | II+III | Late Mousterian | | 35-54 ky | 694 | *Cervus elaphuhs* + *Capreolus capreolus* | temperate climate with humid conditions and woodland covering |
| | RF-Fumane | A4 | Late Mousterian | 45-44 ky BP | - | 484 | *Cervus elaphuhs* + *Capra ibex* | cold-temperate climate with alpine setting and open environments |
| | RB-Broion | 1e+1f+1g | Uluzzian | 38 ky BP | - | 59 | *Sus scrofa* | cold-temperate climate with humid woodlands |
| | RF-Fumane | A3 | Uluzzian | 44-42 ky BP | - | 452 | *Cervus elaphuhs* + *Capra ibex* | cold-temperate climate with alpine setting and open environments |
| | RF-Fumane | A2-A2R | Protoaurignacian | 40-34 ky BP | - | 795 | *Capra ibex* | cold climate with steppic enviroments |
| Southern Italy – Tyrrhenian | Cala | R | Late Mousterian | - | - | - | - | - |
| | CTC-Castelcivita | 32-21 | Late Mousterian | 46-42 ky BP | - | 453 | *Dama dama* + *Cervus elaphus* | temperate woodland covering |
| | CTC-Castelcivita | 20-18lower | Late Mousterian | | - | | *Rupicapra* sp. + *Cervus elaphus* | woodland covering and increasing in humidity |
| | CTC-Castelcivita | 18upper-13 | Uluzzian | - | - | 134 | *Capreolus capreolus* + *Rupicapra* sp. | temperate climate with more dispersed woodlands |
| | CTC-Castelcivita | 12-10 | Uluzzian | 42-40.5 ky BP | - | 110 | *Equus ferus* | cold climate and increased presence of open enviroments |
| | Cala | 14 | Uluzzian | - | - | 347 | *Dama dama* | temperate climate and mediterranean evergreen |
| | CTC-Castelcivita | 10upper-8 | Protoaurignacian | - | - | 33 | *Equus ferus* + *Sus scrofa* | cold climate with woodland covering and open enviroments |
| | CTC-Castelcivita | 7-top sequence | Protoaurignacian | - | - | 60 | *Cervus elaphus* + *Rupicapra* sp. | cold-temperate climate |
| | Cala | 13 | Protoaurignacian | - | - | 230 | | |
| | Cala | 12 | Protoaurignacian | - | - | 428 | *Cervus elaphus* | onset of cold climate with dispersal woodlands |
| | Cala | 11-10 | Protoaurignacian | - | - | 228 | | |
| Southern Italy – Ionian-Adriatic | CAV-Cavallo | FIIIE | Late Mousterian | - | - | 349 | *Bos primigenius* + *Cervus elaphus* | open/forest steppe |
| | CAV-Cavallo | FIIIB-D | Late Mousterian | - | - | 268 | *Dama dama* + *Bos primigenius* | temperate phase |
| | CAV-Cavallo | FIIIA-FI | Late Mousterian | >45 ky BP | - | 253 | *Bos primigenius* + *Cervus elaphus* | semi-arid stage/forest steppe |
| | OSC-Oscurusciuto | 4-13 | Late Mousterian | - | - | 574 | *Bos primigenius* | wooded meadows and open spaces |
| | OSC-Oscurusciuto | 3 | Late Mousterian | - | - | 57 | *Equus ferus* + *Bos primigenius* | semi-arid stage/forest steppe |
| | OSC-Oscurusciuto | 2-29-30-31 | Late Mousterian | - | - | 185 | *Bos primigenius* + *Equus ferus* | semi-arid stage/forest steppe |
| | OSC-Oscurusciuto | 1 | Late Mousterian | 43-42 ky BP | - | 40 | *Bos primigenius* + *Cervus elaphus* | temperate phase |
| | CAV-Cavallo | EIII | Uluzzian | 45-43 ky BP | - | 194 | *Bos primigenius* + *Equus ferus* | cold climate with more dispersed woodlands |

cold
temperate
semi-arid
cold/temperate

**Table 1**
Detailed context table of sites analysed in this work with reference to dominant taxa and to the most represented environmnetal setting recorded at each site.

**Riparo del Broion** is located in the northern part of the Berici eastern slope, at 135 m asl, along a steep slope comprising escarpments, cliffs and remnants of collapsed sinkholes that connects the top of Mount Brosimo (327 m asl) to the marshy and swampy plain (De Stefani et al., 2005; Gurioli et al., 2006; Romandini et al., 2012; Peresani et al., 2019).

Slope-waste clay deposits can be found at the feet of Mount Brosimo. Uluzzian faunal assemblages (levels 1f-1g) show a high richness due to the different environments of the surroundings. Alongside the presence of marmot, hare, chamois, ibex, bison and possibly aurochs, the number of red deer and roe deer bones as well as the abundance of wild boar remains indicate the existence of humid woodlands located in the alluvial plain to the east of Mount Brosimo (Peresani et al., 2019).

**Grotta Maggiore di San Bernardino** opens on the eastern slope of the Berici karst plateau 135 m asl, to the west of the alluvial plain of the Bacchiglione River. Eight lithological units compose a Middle-Late Pleistocene stratigraphic sequence (Cassoli and Tagliacozzo, 1994b; Peresani, 2001). The majority of the total faunal remains found at the site (78%) belongs to ungulates, although the frequency of ungulate remains varies between stratigraphic units (Table 1). Units II + III, associated to late Mousterian frequentation, is the only layer to have undergone a detailed zooarchaeological study. Its assemblage suggests the presence of humid climatic conditions, the expansion of woodlands (Cassoli and Tagliacozzo, 1994b; Peresani, 2011; López-García et al., 2017; Romandini et al., 2018b; Terlato et al., 2019).

**Grotta del Rio Secco** is located in a stream gorge at 580 m asl on the Pradis Plateau in the eastern part of the Carnic Pre-Alps (Fig. 1 and Table 1), an orographic system dissected by N–S and W-E valleys separating mountains with peaks of 2000–2300 m asl. The site is a flat and wide south-facing shelter, with a gallery completely filled with sediments. The outer area of the shelter presents with a heap of large boulders collapsed from the original, larger roof. Human occupation has been dated (Tables 1 and A.1) to the Late Mousterian (layers 5top, 7, 5, and 8) and to the Gravettian (layers 6 and 4) (Peresani et al., 2014; Talamo et al., 2014). In layers 7 and 8 archaeologists found evidence of the use of fire and of an intensive exploitation of carnivores (*Ursus arctos*, *Ursus spelaeus*, mustelids, and canids), which are more numerous than ungulates (Peresani et al., 2014; Romandini et al., 2018a). Although bird remains are rare, a terminal pedal phalanx of a golden eagle with anthropic cut marks on the proximal articular facet was recovered from layer 7 (Romandini et al., 2014b).

*2.2. Northwestern (Tyrrhenian) Italy*

The coastal area of this region is characterized by a particular relief pattern with middle-range mountains and a narrow littoral plain between the Mediterranean Sea and the southern Alps (Fig. 1). Faunal assemblages of the region date to between Marine Isotope Stage (MIS) 6 and 3, in agreement with geochronological, pollen and cultural data (Valensi and Psathi, 2004; Bertola et al., 2013; Romandini, 2017). From a general point of view, faunal assemblages attributed to the Middle to Upper Palaeolithic transition exhibit high taxonomical richness, reflecting a variety of biotopes such as forest hills, coastal plains, narrow valleys in the hinterland and numerous cliffs. Consistently high values in species richness, in particular for carnivores, were recorded in Liguria during MIS 3 and 2 (Valensi and Psathi, 2004). The most frequent species of ungulates and small mammals point to the extensive presence of forested environments.

A variety of Late Mousterian sites are reported (Fig. 1): Arma delle Manie, Grotta degli Zerbi, Riparo Bombrini, Riparo Mochi, Grotta del Principe in Italy, and Grotte de l'Observatoire in the Principality of Monaco. The arrival of modern humans in the region is associated to a marked change in the archaeological record (Negrino and Riel-Salvatore, 2018; Riel-Salvatore and Negrino, 2018a). At present, Protoaurignacian evidence has been uncovered at Riparo Mochi (Alhaique, 2000; Kuhn and Stiner, 1998; Douka et al., 2012; Grimaldi et al., 2014), Riparo Bombrini (Bertola et al., 2013; Holt et al., 2019; Negrino and Riel-Salvatore, 2018; Riel-Salvatore et al., 2013; Riel-Salvatore and Negrino, 2018a, 2018b), Arma degli Zerbi and Grotte de l'Observatoire (Rossoni-Notter et al., 2016; Onoratini, 2004; Onoratini and Simon, 2006; Porraz et al., 2010; Romandini, 2017).

### *2.3. Southern Italy*

Palaeoecological data for southern Italy come from the Lago Grande di Monticchio record (Monte Vulture, Basilicata). During MIS 3, pollen data associations indicate an alternation between cold/dry steppic vegetation (*Artemisia*-dominated steppe/wooded steppe), related to Greenland Stadials/Heinrich events (GSs/HEs), and an increased range of woody taxa including deciduous *Quercus*, *Abies* and *Fagus* (up to 30–60% of arboreal pollen), referred to Greenland Interstadials (GIs) with a maximum expansion between ca. 55-50 ka (i.e. GI 14) (Allen et al., 1999; Fletcher et al., 2010; Badino et al., this Special Issue). Nevertheless, faunal assemblages coming from MIS3-aged stratigraphic sequences highlight different climatic trends between Tyrrhenian (southwestern) and Ionian (southeastern) contexts (Boscato, 2017) due to an almost persistent moisture availability on the former, mainly generated by the orographic uplift of air charged with moisture from the Tyrrhenian Sea, and to Balkan influence on the latter. The Ionian area is characterized by open environment taxa (e.g. *Bos primigenius*) while the Tyrrhenian one shows an abundance of forest species (Cervidae).

#### *2.3.1. Central - southwestern (Tyrrhenian) Italy*

Southwestern Italy (Tyrrhenian Area – Table 1, Fig. 1 - Area 2) is best represented by **Grotta di Castelcivita** (Salerno). This site is located 94 m asl and is about 20 km far from the modern coastline, in a territory encompassing the valley of the Calore river and the Alburni mountains (m 1742). The archaeological sequence is dated to MIS 3 (Gambassini, 1997) and is about 2.5 m thick. The lowermost portion (layers cgr, gar, lower rsi, spits 32-18lower) contains Late Mousterian deposits and is overlaid by Uluzzian layers (upper rsi, pie, rpi, rsa'', spits 18upper-10lower). The sequence is capped by Protoaurignacian layers (rsa'-gic-ars, spits 10upper-top of sequence), which are sealed by the Campanian Ignimbrite (Giaccio et al., 2017). From a zooarchaeological point of view, a unique aspect of this site is the presence of freshwater fish in all chronological phases (Cassoli and Tagliacozzo, 1997).

**Grotta della Cala** (Marina di Camerota – Salerno) opens close to the present coastline into a steep calcareous cliff which is part of a hilly/ mountain range characterized by plateaus and valleys. The MIS 3 coastline was about 5 km from the cave entrance. The stratigraphic sequence is about 3 m thick and starts from the bottom with Middle Palaeolithic layers in a succession of stalagmites and clastic sediments (Martini et al., 2018). At the entrance of the cave, the Middle Palaeolithic is followed by early Upper Palaeolithic deposits, containing Uluzzian (spit 14) and Protoaurignacian (spits 13-10) evidence (Benini et al., 1997; Boscato et al., 1997). These are covered, after a stratigraphic hiatus, by Gravettian, Epigravettian, Mesolithic and Neo-Eneolithic layers (Palma di Cesnola, 1993).

Beyond these well-documented sites, the only other Uluzzian deposit with faunal assemblages in the region is documented at the Tuscan site of Grotta la Fabbrica (Grosseto; Pitti et al., 1976). Here the abundance of equids points to open environments (less evident in the Protoaurignacian layers). As far as the Late Mousterian is concerned, a similar faunal composition is recorded at Buca della Iena (Lucca; Stiner, 1994). Cervidae are, in contrast, the most abundant family in coeval deposits of Grotta dei Santi (Monte Argentario, Grosseto), suggesting a more humid/temperate climate. In Latium a temperate/humid phase connoted by abundant aurochs and deer remains is recorded at Grotta del Fossellone (Alhaique and Tagliacozzo, 2000) and at Grotta di S. Agostino (Stiner, 1994). A similar trend is found at Grotta Breuil (Alhaique and Tagliacozzo, 2000) where Cervidae are the most abundant in layers 6 and 3a, although ibex remains become more frequent in the latter. In Campania, at Riparo del Poggio (Marina di Camerota), located near Grotta della Cala, faunal assemblages are dominated by Cervidae and are typically linked to temperate climates (fallow deer is the most abundant species; Boscato et al., 2009).

#### *2.3.2. Central - southeastern (Adriatic-Ionian) Italy*

In the Ionian area (Table 1, Fig. 1 - Area 3) **Grotta del Cavallo** opens into the rocky coast of Uluzzo Tower Bay, at the margin of a vast rolling plain. This cave contains a 7- meter thick stratigraphy which has at its bottom a marine conglomerate attributed to MIS 5e. This is overlaid by a Mousterian sequence dated between MIS 5 and MIS 3 (Table 1) (layers N-FI). These layers are followed by an Uluzzian sequence (EIII – DIb; Moroni et al., 2018) sandwiched between two *tephra* layers (Fa and CII) attributed to the Y-6 and the Y-5 (Campanian Ignimbrite) events, respectively (Zanchetta et al., 2018).

**Riparo l'Oscurusciuto** opens inside the ravine of Ginosa (Taranto), to the north of the modern village and about 20 km from the present coastline (Fig. 1). The zooarchaeological data suggest that Neandertal hunters exploited both the main regional environments, i.e. forest steppe located on flat hills and forested area on the humid bottom of the gorge.

The Middle Palaeolithic stratigraphy is 6-m thick. A tephra (US 14) attributed to the Green Tuff of Monte Epomeo (Ischia) and dated to ca. 55 ky seals the surface of a living floor currently under excavation (US 15) (Boscato et al., 2004, 2011; Boscato and Crezzini, 2006, 2012; Boscato and Ronchitelli, 2008). All the cultural assemblages investigated can be referred to MIS 3 and fall in a chronological interval of ca. 12,000 years. Recurrent Levallois is the most abundant lithic production system (Marciani et al., 2016, 2018; Spagnolo et al., 2016, 2018).

In Molise (Adriatic area) Grotta Reali (Rocchetta a Volturno) yielded Late Mousterian assemblages mostly consisting of Cervidae that can be linked to cold and humid climatic conditions (Sala et al., 2012).

## **3. Materials and methods**

Of all the archaeological contexts mentioned in the introduction, the present research only focuses on the 9 ones that present with quantitative evidence on the distribution of faunal assemblages in Middle-to-Upper Palaeolithic transition deposits across Italy (>50-35 ky, Tables 1 and A.9 and Fig. 1). Sampled archaeological sites were grouped into three geographic areas based on site location and ecological/environmental context: 1) Northeastern Italy (4 sites); 2) Southwestern/ Tyrrhenian Italy (2 sites) and 3) Southeastern/Ionian Italy (3 sites; Fig. 1). New zooarchaeological data for Northwestern Italy are now available from Riparo Bombrini (Pothier Bouchard et al., 2019), while for the southeastern/Ionian area the zooarchaeological analysis from Grotta di Uluzzo C is currently in progress (Fig. 1). Both sites are part of the ERC n. 724046 – SUCCESS project, but they are not included in the present synthesis.

All faunal remains used to compute species abundance based on taxon frequency were uncovered by sieving sediment using 0.5 mm and 1 mm meshes in Northeastern, Southwestern, and Southeastern Italy. Based on currently available evidence, specimens were nonetheless grouped into three size classes in Northeastern Italian contexts (0.1–1 cm, 1–3 cm, >3 cm; Table 2) and into two size classes in Southwestern and Southeastern Italian contexts (1–3 cm, >3 cm; Table 5).

Different sources of surface bone alteration (anthropic cut marks vs. animal tooth marks, trampling, postdepositional and modern modifications generated during excavation) were discriminated drawing criteria outlined in a on well-established body of taphonomic literature (Binford, 1981; Potts and Shipman, 1981; Shipman and Rose, 1984; Blumenschine and Selvaggio, 1988; Capaldo and Blumenschine, 1994; Lyman, 1994; Blumenschine, 1995; Fisher, 1995; Fernández-Jalvo and Andrews, 2016; Duches et al., 2016). The degree of combustion was estimated using the method developed by Stiner et al. (1995) and, in Northeastern Italian contexts, burned and calcined bones were separated from unburned materials.

Faunal remains were attributed to species and genus and, when these were not determinable, to families. Unidentified mammal bones were grouped into three classes based on body size: large (red deer, moose, giant deer, bison, aurochs, horse, lion and bear); medium (alpine ibex, chamois, roe deer, fallow deer, wild boar, wolf, lynx, leopard and hyena); and small (hare, marmot, beaver, mustelids, wild cat and fox). In addition, unidentified specimens from the southern sites were grouped according to anatomical categories such as "skull", "jaw", "teeth", "vertebrae", "ribs"etc. or more general categories such as "epiphysis""diaphysis"and "spongy bones".

As far as northern Italy is concerned, taxonomic and skeletal identification were based on the reference collections stored at the Bioarchaeology Section of the National Prehistoric Ethnographic Museum "Luigi Pigorini" (Lazio Museum Pole, Rome), at the Prehistoric and Anthropological Sciences Section in the Department of Humanities, University of Ferrara and at the Laboratory of Osteoarchaeology and Palaeoanthropology at the Department of Cultural Heritage, University of Bologna (Ravenna). Bone assemblages recovered from the southern Italian sites were compared with the reference collection stored at the Research Unit of Anthropology and Prehistory of the University of Siena. Differences between the Uluzzian layer of Grotta del Cavallo and the Late Mousterian layers at Grotta del Cavallo and Riparo l'Oscurusciuto (i.e. the only layers which displayed no sign of carnivore activity on ungulate bones) were formally assessed for percentages of carpal and tarsal bones, and of phalanges and sesamoides (relative frequencies were based on both total ungulate counts and on the remains of *Bos Primigenius*). In addition, the presence of significant differences was tested for remnant diaphysis, epiphysis, and spongy bones between the same layers. To do so we measured effect size as Cohen's *h* using the function ES.h in the package pwr in R (Champely, 2018), we then measured statistical power using the dedicated pwr.2p2n.test function in the same package, and performed a two-tailed test for equality in proportions between the chosen layers (with continuity correction for cases in which the number of successes or failures was lower or equal to 5; Tab. A.13 – A.17). We also tested the hypothesis of differences in the degree of fragmentation across sites of Northern Italy by running arcsine transformation of proportions fragment-size classes at all sites (1–3 cm, >3 cm; following Morin et al., 2019) and then comparing the distribution of transformed values between Uluzzian and Late Mousterian layers via a two-tailed Mann-Whitney test for independent sample design. As for southern Italy, we once again only focused on Riparo l'Oscurusciuto and Grotta del Cavallo in Southeastern Italy. We tested for significant differences in proportions and also calculated effect size and statistical power to support the obtained results.

Species abundance was quantified using the Number of Identified Specimens (NISP; Grayson, 1984). Notwithstanding its limitations (e.g. inflation of the count of rare parts, lower predictive power when limited to long bones), this particular species estimate offers accuracy and reproducibility at the ratio scale (Morin et al., 2017). The ubiquitous recording of NISP in all the examined contexts made it the best available method to directly compare different sites across the study region. Once NISP of each mammal group or species was obtained for all layers of the 9 sampled archaeological sites across Italy, we grouped layers belonging to the same region (i.e. Northeastern, Southwestern, and Southeastern Italy) and within each region we ordered them into a single diachronic sequence, based on absolute dates (Tables 1 and A.9Table A9) and associated material cultural evidence. Avifaunal remains were comapred only for Fumane and Castelcivita caves for abundance reasons and lack of comparable sequences in any other sites. Relative taxon abundance was calculated in each layer and variability in relative frequency over time was inspected through bar charts, in order to highlight any differences between trends emerging in different regions.

Comparing NISP proportions across different archaeological layers (in the same context or between different contexts), however, presents a number of potential issues. In addition to post-depositional processes, substantial differences in the time of accumulation of different layers may have deleterious effects on the accurate representation of faunal spectra. This process, known as time-averaging, is extremely frequent in geologic and anthropic contexts (Binford, 1981; Kowalewski, 1996; Premo, 2014; Madsen, 2018), and has a direct impact on the reliability of the quantification of abundance, richness, evenness, and diversity in time-averaged samples (Leonard and Jones, 1989). Specifically, the longer the duration of layer formation, the more inflated richness and diversity will be. This makes tracking change over time more problematic and increases the risk of misidentifying inflated counts for actual human choices (i.e. Type I error when testing hypotheses; Premo, 2014; Madsen, 2018). The presence of differential accumulation rates, palimpsests, and taphonomic processes therefore complicates any attempt at quantifying the effective temporal scale of individual layers solely based on stratigraphy.

| US - Levels | Technocomplex | 0.1 - 1 cm | % | 1 - 3 cm | % | > 3 cm | % | TOTAL Rem. | Burn.+Calc. | % | Burned | % | Calcined | % |
|---|---|---|---|---|---|---|---|---|---|---|---|---|---|---|
| RF A2-A2R | PA | 13042 | 65.8 | 6280 | 31.7 | 507 | 2.6 | 19829 | 7861 | 40 | | | | |
| RF A3 | UL | 7831 | 46.1 | 8231 | 48.4 | 927 | 5.5 | 16989 | 4723 | 28 | 2840 | 60.1 | 1883 | 39.9 |
| RB 1e+1f+1g | UL | 33199 | 88.8 | 3748 | 10 | 443 | 1.2 | 37390 | 18464 | 49 | 15595 | 84.5 | 2869 | 15.5 |
| RF A4 | LM | 9770 | 49 | 9287 | 46.5 | 898 | 4.5 | 19955 | 7321 | 37 | 5187 | 70.9 | 2134 | 29.1 |
| SB II+III | LM | 2744 | 29.8 | 5337 | 57.9 | 1136 | 12.3 | 9217 | 5431 | 59 | 4747 | 87.4 | 684 | 12.6 |
| RS 5top+7 | LM | 43 | 8.2 | 47 | 9 | 434 | 82.8 | 524 | 693 | 8 | 42 | 97.7 | 1 | 2.3 |
| RF A5/A5+A6 | LM | 35342 | 52.7 | 29767 | 44.4 | 1974 | 2.9 | 67083 | 38255 | 57 | 30442 | 79.6 | 7813 | 20.4 |
| RF A6 | LM | 62692 | 56.5 | 43944 | 39.6 | 4408 | 4 | 111044 | 53413 | 48 | 46854 | 87.7 | 6559 | 12.3 |
| RF A9 | LM | 78119 | 69.8 | 30763 | 27.5 | 2959 | 2.6 | 111841 | 54411 | 49 | 50398 | 92.6 | 4013 | 7.4 |
| RS 5+8 | LM | 2307 | 53.6 | 1538 | 35.8 | 456 | 10.6 | 4301 | 43 | 16 | 671 | 96.8 | 22 | 3.2 |

**Tab. 2**
Different size classes of mammals bones and burned remains (with relative %) identified in the MP/UP transitional contexts of Northern Italy (see Fig. 1 - Area 1). RF = Grotta di Fumane; RB = Riparo del Broion; SB = Grotta di San Bernardino; RS = Grotta del Rio Secco.

In addition, inference made by comparing NISP proportions is hampered by the limitations of closed datasets (Lyman, 2008; Orton et al., 2017). Species relative frequencies are by definition computed over the total number of collected remains and their sum is bound to be equal to 1. No relative frequency is free to vary over time without affecting or being affected by change in the frequency of another class, i.e. the relative abundance of one particular taxon will always be negatively correlated to the relative abundance of another taxon. Interpreting such increases and decreases as the effect of some independent mechanism (e.g. environmental change, cultural selection) is therefore not always straightforward.

In order to overcome the limitations mentioned above while providing support for the trends that might emerge from relative taxonomic abundance analysis across the time-ordered layers of different sites, we also built long-term time-series of zooarchaeological data documented at Grotta di Fumane (Northeastern Italy) that can directly be compared against independent sources of information (e.g. palaeoclimatic models, palinological and palaeoenvironmental data), and across mismatched and differentially overlapping contexts. Grotta di Fumane was chosen as a case study because it offers the longest and best-dated sequence among all the available sites.

More specifically, we computed aoristic sums (i.e. the sum of the probability of existence of all events for a given temporal interval) of taxon abundance to obtain estimates of taxon frequency based on absolute radiocarbon dates. Aoristic analysis has been already employed in a few archaeological and zooarchaeological studies (Ratcliffe, 2000; Johnson, 2004; Crema, 2012; Bevan et al., 2013; Orton et al., 2017), although the method is still generally rarely used and, to the best of our knowledge, it has never been applied to palaeolithic contexts. This approach consists of: a) assigning a start and end date to each archaeological layer from which fossil fragments had been retrieved; b) dividing the entire time span of the study period into temporal bins of fixed width; c) based on the start and end dates of the relevant layer, and drawing on Laplace's principle of insufficient reason (see Crema, 2012; Orton et al., 2017 for a detailed discussion), dividing the total probability mass of each deposition event/fragment (equal to 1) across the $t$ temporal bins comprised in the date interval of the layer. Each deposition event therefore exhibits a uniform probability of existence at each bin calculated as $1/t$; d) summing all the probabilities falling in the same bin, and repeating the same operation for the entire study period.

The result is an estimate of species frequency distribution which incorporates all the temporal uncertainty embedded in the data. Better dating leads to shorter temporal intervals for each deposition event, that in turn allows researchers to assign a higher probability of existence at each temporal bin. As a consequence, worse dating leads to higher dispersion in the probability of existence, i.e. to stable time series which do not show clear evidence of increase or decrease as an artefact due to lack of resolution. In the present work, we first set the temporal limits for each layer at Grotta di Fumane. When start and end dates were already available from the literature (as in the case of layers A9) these intervals were directly taken (Table A.9Table A 9). As far as all the remaining layers are concerned (A6, A5/A5+A6, A4, A3, and A2), start and end dates were calculated in OxCal 4.3 (Bronk Ramsey, 2009) as the median of the 68.2% interval taken from the posterior probability distribution of already published layer boundaries (Higham et al., 2014). This particular model was chosen to fully exploit the potential of aoristic analysis and considering that at this site Uluzzian and final Mousterian are reported as temporally indistinguishable (Douka et al., 2014).

Raw NISP counts were then used to compute aoristic sums of each taxon across 50-year bins through the function *aorist* in the package *archSeries* in R version 3.4.4 (Orton, 2017, R Core Team 2018). To avoid generating artifacts due to empty bins at the interval 41600-41100 cal BP, 10 years were added to the median date for the end boundary of level A3. Taxon-specific aoristic values were then summed and used to calculate estimates of taxon relative frequencies. To further ascertain the presence of absolute shifts in estimated frequency, we also plotted the aoristic sum of ungulates. In this case, absolute frequency estimates were compared against 95% confidence envelopes generated through Monte-Carlo simulation (n. iterations = 5000) as well as against a dummy model generated assuming a uniform frequency distribution following Crema (2012) and Orton (Orton et al., 2017). Both the empirical and dummy simulations were computed using the function *date.simulate* in the package *archSeries*. Using the same function, rates of change were also computed for ungulate families. The aim was to assess whether there were temporal bins exhibiting shifts in the abundance of families compared to other bins. Following Crema (2012) and Orton (2017), rates of change were examined by observing (in this case through boxplots) the distribution of simulated standardised differences between each chronological bin and the preceding one. Temporal intervals with median and interquartile range falling above the zero line (suggesting stability or absence of change) were interpreted as a sign of increase, while boxes falling under the zero line were interpreted as instances of decrease. Such distributions were compared against the 95% confidence envelopes of the null model based on the aoristic sum of carnivores, which provides a null expectation independent from palaeoenvironmental change.

## 4. Results

Northeastern Italian contexts yielded a total of 323,964 remains (NISP = 9044) while for Southern Italy as a whole 33,340 remains were documented (NISP = 2351). From a zooarchaeological point of view, Late Mousterian layers have been investigated more intensively than later ones in both regions. Despite the difference in absolute counts, the proportion of mammal orders and classes is roughly the same across all contexts (Fig. 2), with ungulates being the most abundant category followed by carnivores, birds (at Grotta di Fumane and Castelcivita), and rodents, in decreasing order of importance.

Uluzzian layers exhibit an increase in the relative abundance of carnivore and bird remains, matched by a considerably lower number of remains attributed to large rodents (e.g. marmot and beaver) and lagomorphs (Fig. 2). Protoaurignacian phases invert this trend, with an appreciable decrease in the number of carnivore and bird remains.

### *4.1. Mammals*

Despite the specificities that may bias the abundance of faunal remains in each of the examined contexts (e.g., Grotta del Rio Secco being consistently used by bears which, in turn, were routinely exploited by Neandertals; Romandini et al., 2018b), most of Late Mousterian levels and layers in Northeastern Italy show an increase in the prevalence of cervidae, followed by a decrease of *Cervus elaphus* and *Capreolus capreolus* matched by a gradual increase, in the Uluzzian and Protoaurignacian, in caprinae, especially *Capra ibex* and *Rupicapra rupicapra (*Fig. 3 and Table A.1*)*. This change over time in the relative abundance between cervidae and caprinae may hint at a shift from a temperate climate characterised by forests and meadows to an alpine setting with open environments.

The archaeological sites are located in a region that included habitats suitable for bovinae, ranging from dense forests with wetlan and small streams more attractive to *Bos primigenius*, to hilly grasslands and plains, populated by bison. However, bovids are generally less abundant than the previous families, and their presence remains roughly constant across the entire study sequence.

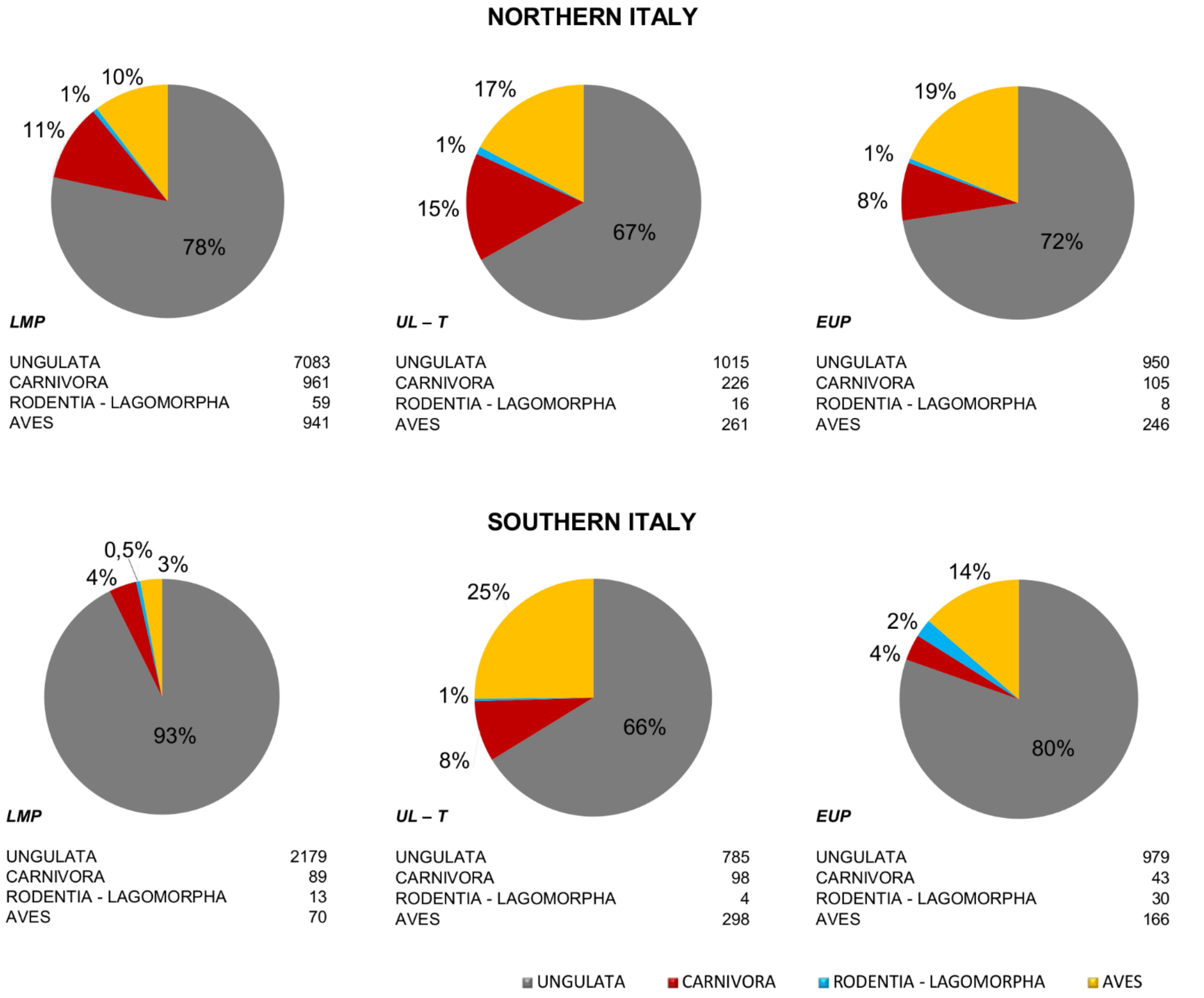

**Fig. 2.** Comparison of the relative frequency of the NISP of Ungulata, Carnivora, Rodentia-Lagomorpha and birds (the latter only for Grotta di Fumane and Castelcivita) documented in all sampled sites (Fig. 1) and divided by macro-geographical area and cultural phase: LM = Late Mousterian; UL = Uluzzian; PA = Protoaurignacian.

Moose (*Alces alces*) and giant deer (*Megaloceros giganteus*) are less frequent and well attested in Mousterian and Uluzzian layers. Their presence suggests – during this period – the existence of humid woodlands near the sampled archaeological sites. Wild boar is rarer yet, being present anecdotally in the Late Mousterian at Grotta di Fumane, while it is more abundant at lower elevations (Grotta di San Bernardino, Mousterian Units II + III; Riparo del Broion, Uluzzian layers 1e+1f+1g). The presence of woolly rhinoceros (*Coelodonta antiquitatis*) in the Uluzzian layer A3 at Grotta di Fumane and of *Stephanorhinus* sp. at Grotta di San Bernardino indicates decreasing temperatures and presence of cold arid conditions.

In this region, carnivores are quantitatively more represented in Late Mousterian and Uluzzian assemblages, while their frequency steadily decreases in Protoaurignacian layers (Fig. 2). Nevertheless, variety of carnivores taxa increase beginning with the Uluzzian (Romandini et al., 2018a), and the presence of wolverine (*Gulo gulo*), ermine (*Mustela erminea*), and arctic fox (*Alopex* cfr. *lagopus*) further supports the onset of colder and arid climate conditions during the MP-UP transition (Fig. 4 and Table A.2). Rodents and lagomorphs (Table A.3) are represented by beaver and marmot, already present in Late Mousterian assemblages, and by lagomorphs in the Uluzzian and Protoaurignacian (Romandini et al., 2018a). Upper Palaeolithic contexts also yielded remains of *Lepus* cfr. *timidus,* further sup porting the diffusion of increasingly colder environments in the latest phase of the studied sequences.

In Southwestern/Tyrrhenian Italy, the Late Mousterian sequence at Grotta di Castelcivita (spits 32-24) yielded a conspicuous amount of cervidae fragments (*Cervus elaphus*, *Dama Dama*, *Capreolus capreolus*); fallow deer in particular is the most abundant species (Fig. 5 and Table A.4). Later on, in spits 23-18 lower, there is an increase in the abundance of red and roe deer and of chamois (*Rupicapra* sp.), correlated to an increase in humidity (Masini and Abbazzi, 1997). The beginning of the Uluzzian sequence (spits 18 upper-15) is characterised by higher frequencies of horse (*Equus ferus*) and large bovids (*Bison priscus* and *Bos/ Bison*) suggesting the occurrence of colder climates and sparse woodland. In the following Uluzzian layers (spits 14-10 lower), an additional increase in the occurrence of equids and a decrease in the frequency of fallow deer suggest more open environments. The Early Protoaurignacian (spits 10 upper – 8 lower) shows comparable environmental conditions, while spits 8upper-7 can be linked to an increase in woodland cover as suggested by the higher presence of deer and the decrease in the frequency of horse (Fig. 5 and Table A.4), followed by cold-temperate phases (spit 6) (Masini and Abbazzi, 1997). The anthracological evidence supports the climatic and ecological trend inferred from zooarchaeological remains (Castelletti and Maspero, 1997).

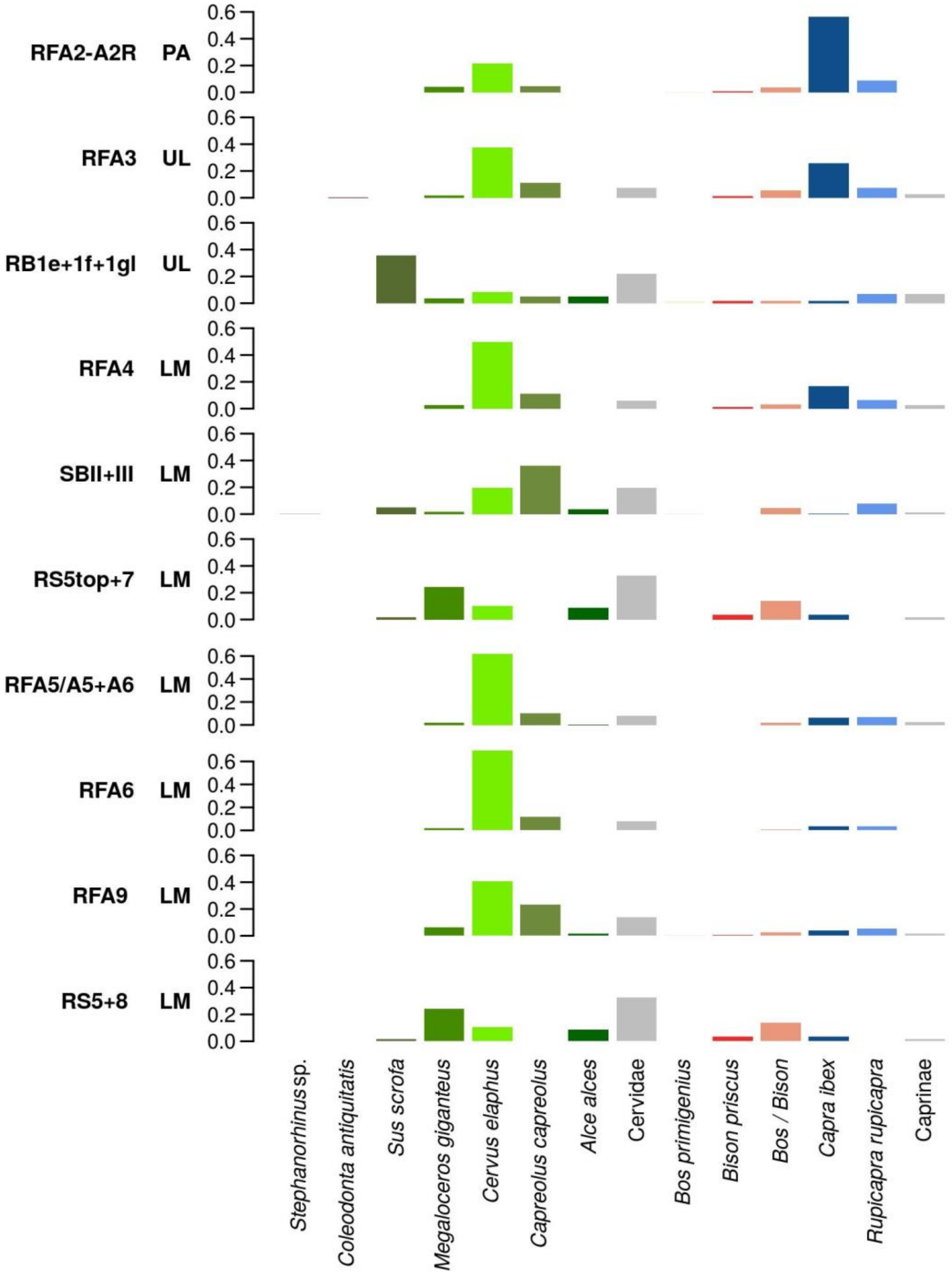

**Fig. 3.** Bar charts showing the relative contribution of each ungulate taxon to the total NISP recorded in the different levels and layers sampled in Northern Italy. The contexts are in chronological-cultural order (from bottom to top) based on the archaeological sequence of each site. RF = Grotta di Fumane; RB = Riparo del Broion; SB = Grotta di San Bernardino; RS = Grotta del Rio Secco. LM = Late Mousterian; UL = Uluzzian; PA = Protoaurignacian.

At Grotta della Cala (Marina di Camerota, Salerno), faunal remains from the Uluzzian (spit 14) are characterised by a conspicuous presence of cervidae (representing on the whole 74% of ungulates) and in particular of fallow deer, typical of temperate climates and Mediterranean evergreen forest. In the Protoaurignacian layers (spits 13-10), lower frequencies of fallow deer and higher frequencies of red deer indicate the onset of colder conditions (Fig. 5 and Table A.4). Low frequencies of ungulates linked to open environments/wooded steppe (such as horse, alpine ibex and aurochs) are also recorded (Benini et al., 1997). Cervids account for over 70% of the ungulates recovered in these layers (Boscato et al., 1997).

In the same region, carnivores occur in all phases. Whilst in the Middle Palaeolithic, most of the remains are referable to the spotted hyaena and the leopard, species richness increase in the Uluzzian and in the Protoaurignacian (Table A.5). Rodents and lagomorphs are very rare. The record of Southeastern/Ionian Italy, on the other hand, is based on the sequences uncovered at Riparo l' Oscurusciuto (Ginosa - Taranto) and Grotta del Cavallo (Nardò - Lecce) (Figs. 1 and 6 and Tables A.6). At Riparo l'Oscurusciuto, layers 13:4 are characterised by the substantial presence of *Bos primigenius*, counterbalanced by low frequencies of horse, rhinoceros and caprinae, and by anecdotal frequencies of cervidae (especially fallow deer), all of which hints at an environment characterised by wooded meadows and open spaces (Fig. 6).

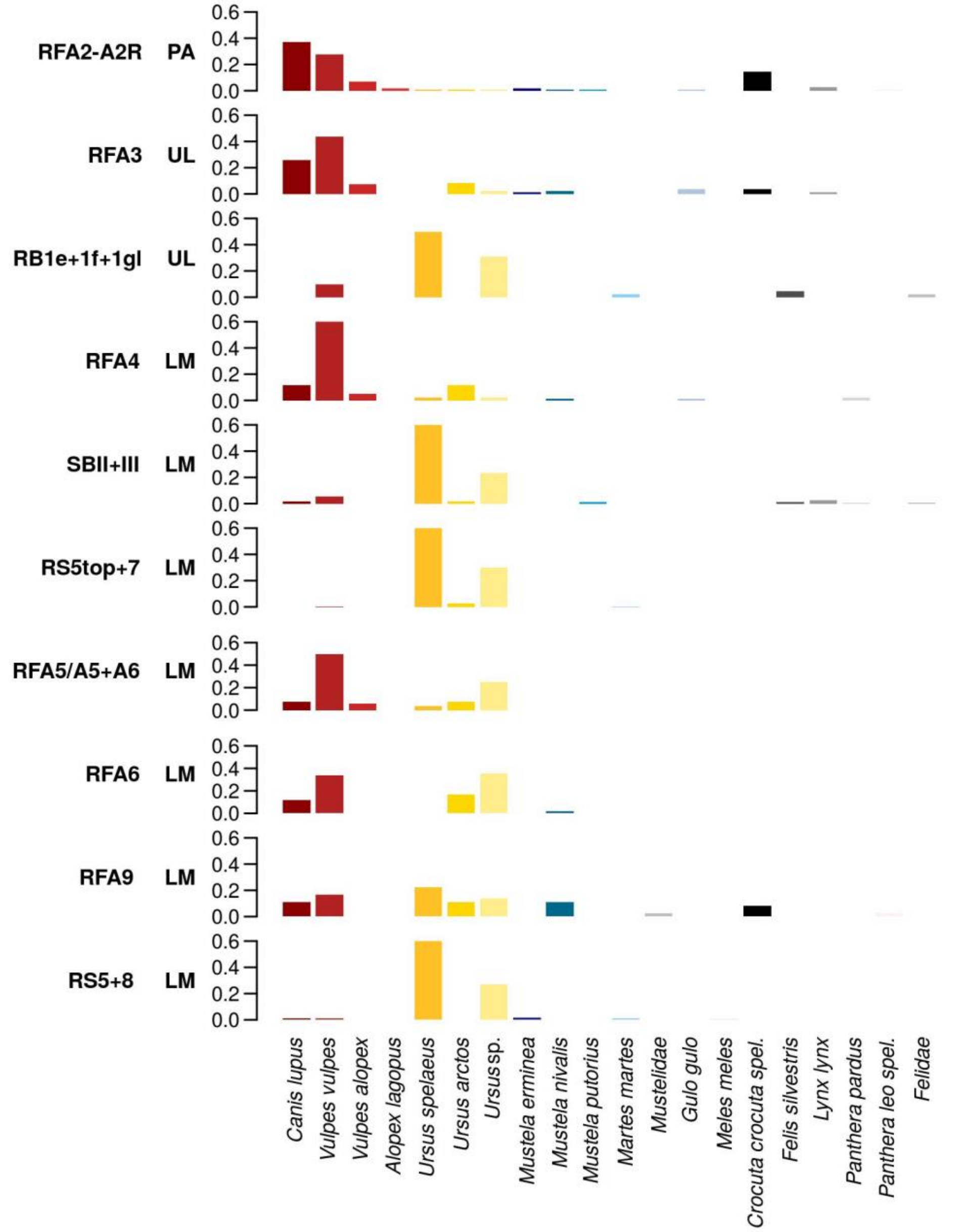

**Fig. 4.** Bar charts showing the relative contribution of each carnivore taxon to the total NISP recorded in the different levels and layers sampled in Northern Italy. The contexts are in chronological-cultural order (from bottom to top) based on the archaeological sequence of each site. RF = Grotta di Fumane; RB = Riparo del Broion; SB = Grotta di San Bernardino; RS = Grotta del Rio Secco. LM = Late Mousterian; UL = Uluzzian; PA = Protoaurignacian.

Aurochs is less frequent in SU 3, while in the same unit, deer is more abundant, the presence of rhinoceros can be inferred by tooth fragments, and horse becomes the most represented ungulate (Table A.6). At the end of the sequence (SU2-1), aurochs is once again the most abundant ungulate, while the increase in abundance of fallow deer suggests the onset of a temperate phase (Boscato and Crezzini, 2012).

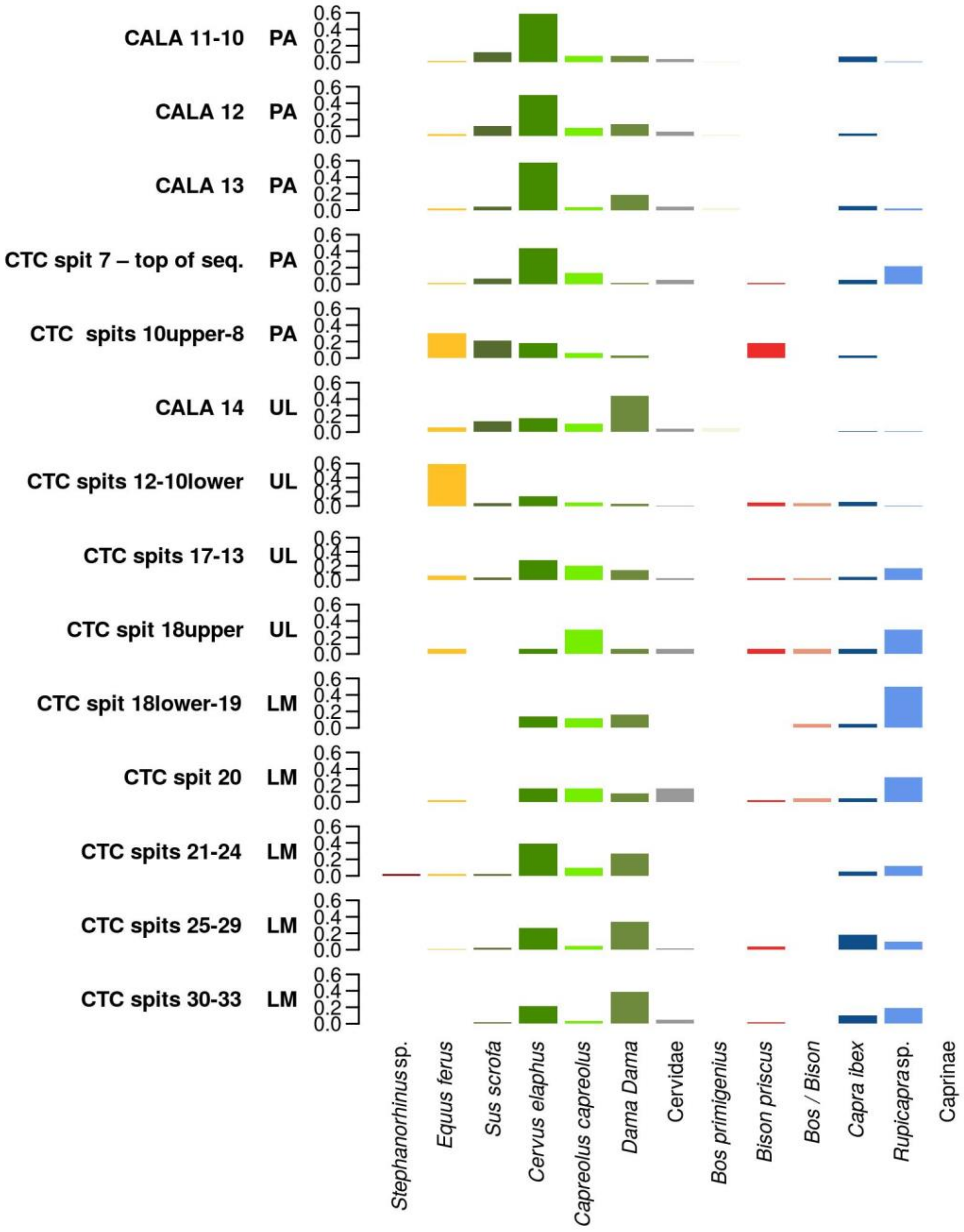

**Fig. 5.** Bar charts showing the relative contribution of each ungulate taxon to the total NISP recorded in the different levels and layers sampled in Southwestern Italy (Tyrrhenian Area). Contexts are presented in chronological-cultural order (from bottom to top) based on the archaeological sequence of each site. CTC = Grotta di Castelcivita; CALA = Grotta della Cala. LM = Late Mousterian; UL = Uluzzian; PA = Protoaurignacian

The Late Mousterian sequence at Grotta del Cavallo (layers FIII-FI) also yields evidence of the climatic fluctuations known for MIS 3 (Table A.6), which agrees with the sequence described for Riparo l'Oscurusciuto. An initial phase characterised by open/steppic forests indicated by the dominance of aurochs remains (layer FIIIE) is followed, in layers FIIID-FIIIB, by a more temperate phase (as suggested by higher frequency of fallow deer) and by a third, more arid stage in layers FIII-FI associated with the presence of aurochs and horse (Sarti et al., 2000, 2002) (Table A.6). The lowermost Uluzzian level EIII5 suggests, in contrast, a shift to an increasingly colder climate with more dispersed woodlands, indicated by the absence of fallow deer and by the increased presence of horses (Table A.6; Boscato and Crezzini, 2012).

With the only exception of red fox which has been found in the Late Mousterian of Grotta del Cavallo, carnivores, rodents and lagomorphs are almost absent in the assemblages of Ionic area (Table A.7).

### 4.2. Avifaunal remains

Substantial evidence on the exploitation of avifauna was documented for Grotta di Fumane and Grotta di Castelcivita (Cassoli and Tagliacozzo, 1994a, 1997; Masini and Abbazzi, 1997; Gala and Tagliacozzo, 2005; Peresani et al., 2011a; Romandini, 2012; Tagliacozzo et al., 2013; Romandini et al., 2016a, b; Gala et al., 2018; Fiore et al., 2004, 2016, 2019). The bird species identified at Grotta di Fumane belong to the extant Italian avifauna (Table A.8) with the exception of the willow grouse (*L.* cf. *lagopus*), a boreal species which has historically never been observed in Italy. The parrot crossbill (*L. pytyopsittacus*) is another boreal species currently found in Northern Europe and consid ered a vagrant species in Italy (Brichetti and Fracasso, 2015). Taxa linked to open and rocky environments are the most abundant (Fig. 7). More specifically, bearded vulture (*G. barbatus*), golden eagle (*A. chrysaetos*), red-billed and yellow-billed chough (*P. pyrrhocorax* and *P. graculus*), common raven (*C. corax*), Eurasian crag martin (*P. rupestris*) and white-winged snowfinch (*M. nivalis*) indicate the presence of rocky cliffs, while treeless terrain with rocky outcrops is indicated by rock partridge (*A. graeca*) and rock ptarmigan (*L. muta*). On the other hand, the presence of wooded areas in the surroundings of the cave is indicated by the black grouse (*L. tetrix*), stock dove (*C. oenas*), common woodpigeon (*C. palumbus*), Boreal owl (*A. funereus*), tawny owl (*S. aluco*), white-backed woodpecker (*D. leucotos,* currently reduced to small populations in the Central Apennines but once more widespread; Pavia, 1999; Brichetti and Fracasso, 2007) Eurasian jay the northern nutcracker (*N. caryocatactes*) and parrot crossbill (*L. pytyopsittacus*) also point to the presence of coniferous forests. Additionally, open grasslands and wet meadows are indicated by common quail (*C. coturnix*), grey partridge (*P. perdix*), corncrake (*C. crex*, which was breeding in the surroundings of the cave, as attested by the finding of juvenile bones) and northern lapwing (*V. vanellus*), while willow grouse (*L.* cf. *lagopus*) is an indicative of tundra-like open areas such as moors and peatlands. Finally, the presence of two duck species (*A. platyrhynchos* and *A.* cf. *crecca*) and two Rallidae (*R. aquaticus* and cf. *G. chloropus*) suggests nearby wetlands or slow-flowing water courses (Cramp, 1998).

As a whole the bird assemblage at Fumane points to an Alpine ecological setting with forests and open areas. Several identified bird species (rock ptarmigan, black grouse, Boreal owl, bearded vulture, white-backed woodpecker, red and yellow-billed chough, Northern nutcracker and white-winged snowfinch) currently live in Italy at considerably higher altitudes than Fumane. The presence of their fossil remains at 350 m asl suggests the downward shifting of the vegetational zones during MIS 3 due to a decrease in climate value parameters.

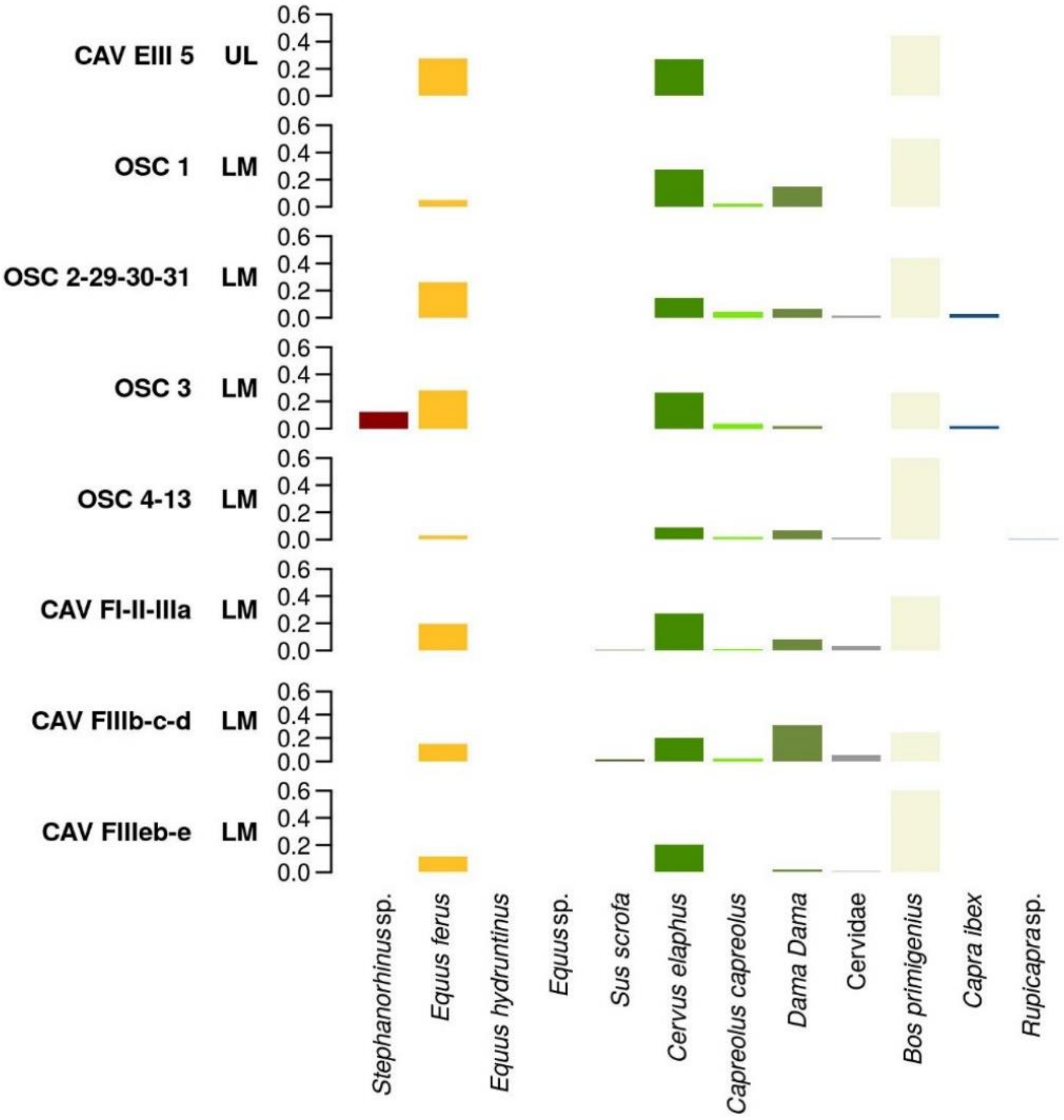

***Fig. 6.*** Bar charts showing the relative contribution of each ungulate taxon to the total NISP recorded in the different levels and layers sampled in Southeastern Italy (Ionian-Adriatic area). Contexts are presented in chronological-cultural order (from bottom to top) based on the archaeological sequence of each site. CAV = Grotta del Cavallo; OSC = Riparo l'Oscurusciuto. LM = Late Mousterian; UL = Uluzzian; PA = Protoaurignacian.

The presence of remains probably belonging to willow grouse in layer A6 and to parrot crossbill in layers A1+A2 during two of the harsher climatic phases (Heinrich Event 5 and Heinrich Event 4 respectively) (López-García et al., 2015) might be an example of two boreal species seeking a *refugium* in Mediterranean Europe (Tyrberg, 1991; Carrera et al., 2018a,b).

The relative frequency of species related to forest, open, rocky and water environments calculated for each layer (Fig. 7) suggests the presence of temperate conditions for layer A9, followed by a colder climate in A6. The species linked to open environments decline sharply in layer A4 (attributed to the GI12 interstadial) before increasing anew in layer A3, marking the beginning of Heinrich Event 4 that lasts until the end of the sequence (López-García et al., 2015). Heinrich Event 4, however, did not prevent the persistence of open forests, as attested by the bird taxa found in A1+A2 and A3. The apparent increase of forest bird taxa in the layers corresponding to Heinrich Event 4 could be explained by a switch from anthropic to natural accumulation

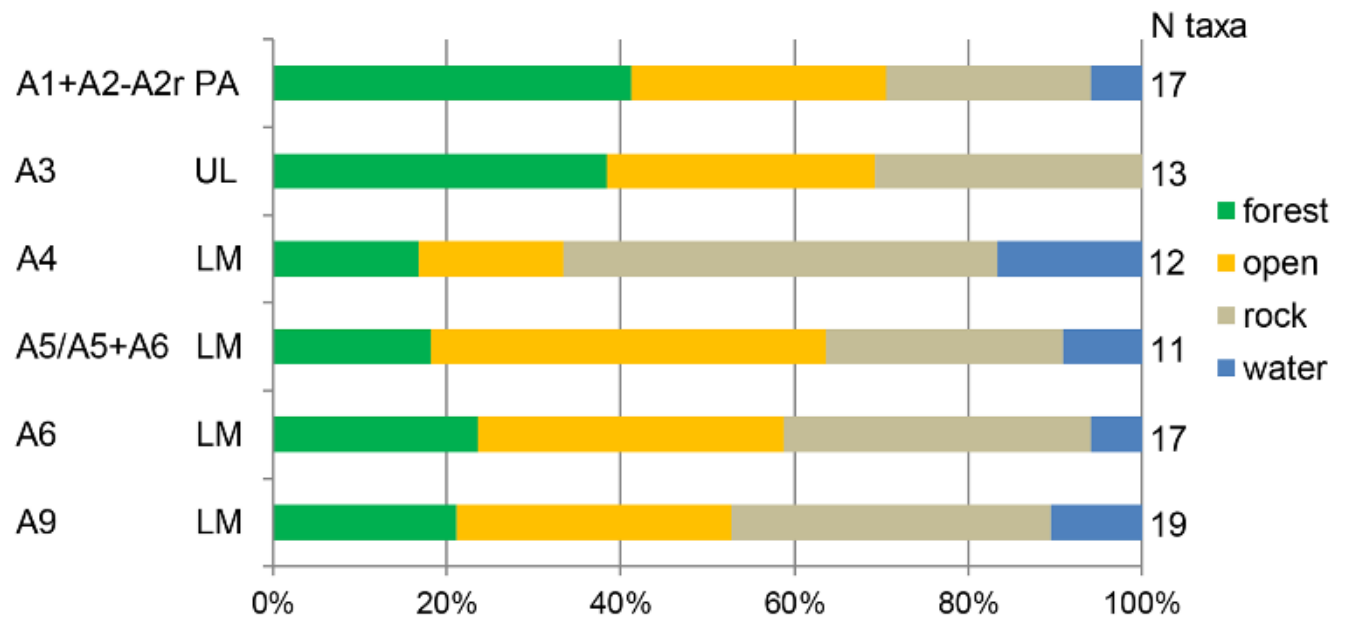

***Fig. 7.*** % share of the bird taxa of different environments (see legend) in the various layers of Grotta di Fumane, calculated on the total number of bird taxa of each layer. LM = Late Mousterian; UL = Uluzzian; PA = Protoaurignacian.

in the Protoaurignacian, as suggested by the lack of anthropic marks on bird bones.

Turning to Grotta di Castelcivita, while the bird remains have been the object of a recent taphonomic revision (Fiore et al., 2019), the data discussed here are drawn from Cassoli and Tagliacozzo (1997). All identified taxa belong to the extant Italian avifauna (Table A.8 and Fig. 8, Cassoli and Tagliacozzo, 1997; Brichetti and Fracasso, 2015; Gala et al., 2018). In terms of NISP, the vast majority are from open and rocky environments. The presence of extensive wetlands and marshes near the cave is suggested by several duck, wader and gull species (*A. nyroca*, *S. querquedula*, *M. strepera*, *M. penelope*, *A. platyrhynchos*, *A. crecca*, *P. squatarola*, *N. phaeopus*, *L. limosa*, *A. interpres*, *C. pugnax*, *G. media*, *L. ridibundus*). The red-billed chough (*P. pyrrhocorax*), yellow-billed chough (*P. graculus*) and Eurasian crag martin (*P. rupestris*) are indicative of rocky cliffs, while rock partridge (*A. graeca*) attests to the presence of treeless rocky terrain. Open areas such as grasslands, steppe and shrublands were also present, as indicated by the common quail (*C. coturnix*), grey partridge (*P. perdix*), Eurasian thick-knee (*B. oedicnemus*), and little owl (*A. noctua*).

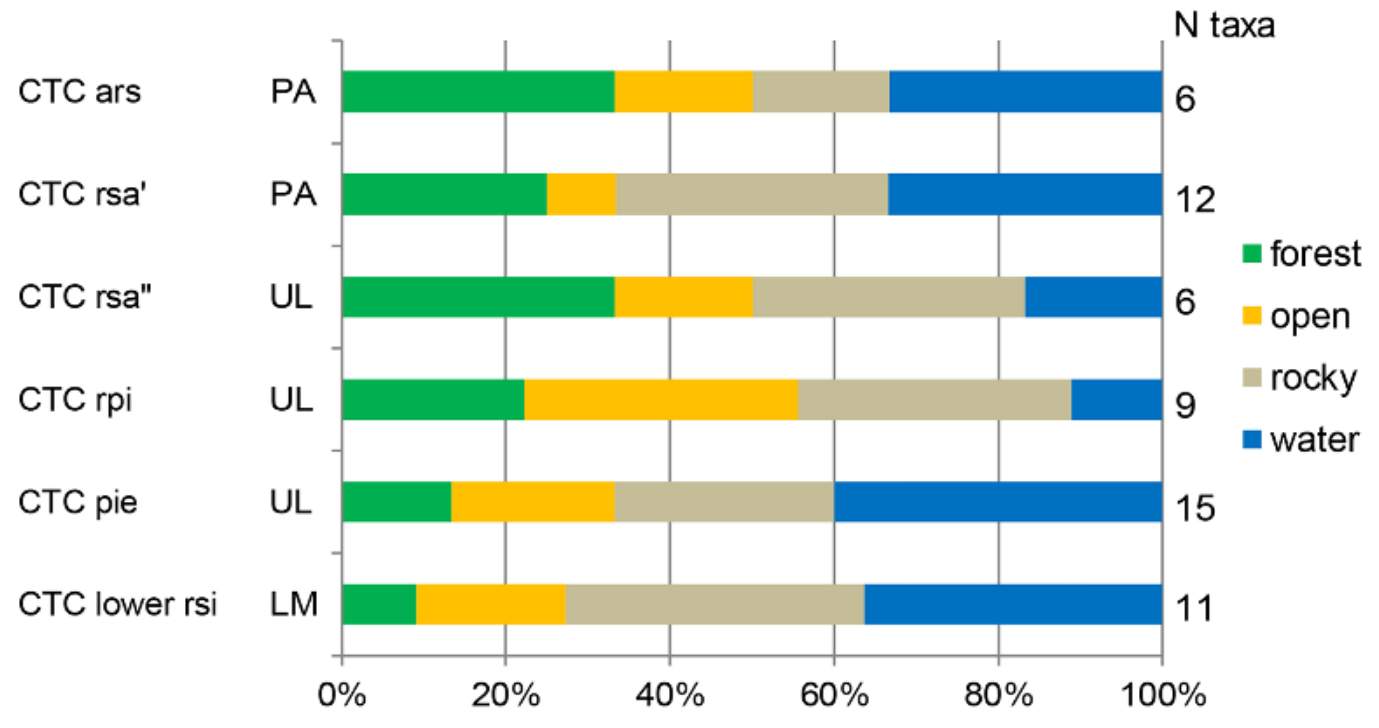

**Fig. 8.** % share of the bird taxa of different environments (see legend) in the various layers of Grotta di Castelcivita, calculated on the total number of bird taxa of each layer. LM = Late Mousterian; UL = Uluzzian; PA = Protoaurignacian.

The stock dove (*C. oenas*) and tawny owl (*S. aluco*) are associated with wooded areas, while the presence of the northern nutcracker (*N. caryocatactes*) suggests the presence of coniferous forests (Cramp, 1998) and confirms that, in the past, this species was distributed across a much broader area than today (Gala and Tagliacozzo, 2010; Brichetti and Fracasso, 2011).

The surroundings of Castelcivita were therefore characterised during MIS 3 by extensive wetlands in plain areas, and by drier environments (such as grasslands, bare terrains and cliffs) intermingled with conifer or mixed forests at higher elevations. The presence at about 100 m asl of species that currently live at higher altitudes (rock partridge, choughs and northern nutcracker), suggests colder and harsher conditions during the whole sequence. In the Uluzzian layer rpi, the number of bird taxa related to open environments increases and point to an expansion of grasslands linked to colder and more arid conditions possibly corresponding to the beginning of Heinrich Event 4 (or a preceding stadial), based on currently available dates (Fig. 8). In the Protoaurignacian layers, the riparian taxa slightly increase, as do those of forest environments in layer ars, probably indicating a climatic amelioration at the top of the sequence (Cassoli and Tagliacozzo, 1997; Gala et al., 2018). All phases provided evidence of human exploitation of at least some bird species (Fiore et al., 2019).

### *4.3. Aoristic analysis*

As far as estimated relative frequencies of ungulates are concerned, the resulting graph exhibits trends of change over time (Fig. 9 B). The estimates of *Capreolus* relative abundance are high at 47.6 ky while they decrease after ~45 ky. After ~45 ky the estimated remains of *Cervus elaphus* start becoming more abundant than in previous bins and reach their maximum between 45-44 ky. In bins following that date the presence of red deer starts declining in favour of *Capra ibex* and *Rupicapra rupicapra*. In the same time interval *Bos* and *Bison* show a quick unimodal trend. The apparent stability that emerges after ~40.5 ky it is due to the assumed uniform probability distribution in the absence of additional information on layer chronology. This trend, which can be noticed for all taxa, could be an artefact of uneven chronological sampling, and underscores the great temporal uncertainty associated with Protoaurignacian assemblages.

The distribution of relative carnivore estimates (Fig. 10) shows a marked increase in the presence of Ursidae between ~45 and ~43.6 ky. The relative frequency of wolves becomes higher than that of Ursidae between ~44.1 and 43.6 ky and becomes the highest value from 41.1 ky onwards. Foxes become most frequent when Ursidae increase and exhibit a inverse trend to that of wolf. From both an environmental and taphonomic point of view, it is interesting to note that after 41.6 ky the estimated relative frequency of *Crocuta crocuta spelaea* is considerably higher than in previous bins, including bins that record its presence between 47.6-46.6 ka.

Birds adapted to rocky environments are the most frequent at Fumane for the entire study sequence (Fig. 11). At 44.1 ky they exhibit a much lower estimate, while the percentage of avifaunal specimens linked to wooded/forested environments is higherthan it was at the beginning of the sequence.

Rates of change based on simulated dates for ungulates (Figs. 9 and 11) add interesting elements and support this emerging scenario. When trends for ungulates are plotted against dummy sets based on estimated absolute frequencies of carnivores (i.e. of species with no particular links to change in environmental conditions), simulated 95% confidence envelopes exceed the expectations of the dummy model confidence area between 47.6-45.6 ky for bovids (Fig. 12 C). Although median values of simulated rates of change for all three ungulate families never emerge from the 95% dummy confidence envelope – suggesting the lack of significant deviations from a null model of deposition which is supposedly not based on environmental change – box-plots consistently point to the same chronological bins as the interesting ones. More specifically, there is evidence of a possible absolute increase in the frequency of all ungulate families between ~45.6-45.1 ky, while a decrease could be hypothesis between 45.1-44.1 ky (see Fig. 12).

Aoristic sum and simulated frequency estimates of all ungulates as a whole (Fig. 9 A) further support the hypothesis that at Grotta di Fumane there was an intensification of the deposition of ungulates between ca 45 and 44 ky, coinciding with higher percentages of red deer in the assemblages. A second moment of more intense process could be identified between 41 and 39 ky, corresponding to higher percentages of *Capra ibex* and *Rupicapra rupicapra*. These trends confirm what emerged through the inspection of relative frequency estimates and of relative taxonomic abundance analysis, and hint at a potential change in environmental and climatic conditions in the region, but also point to a possible change in hunting and subistence strategies, in particular by comparing ungulates, carnivores, and small preys from taxon abundance analysis. on the same portions of other ungulates, many of which were likely used as retouchers (Jéquier et al., 2018; Romandini et al., 2018a).

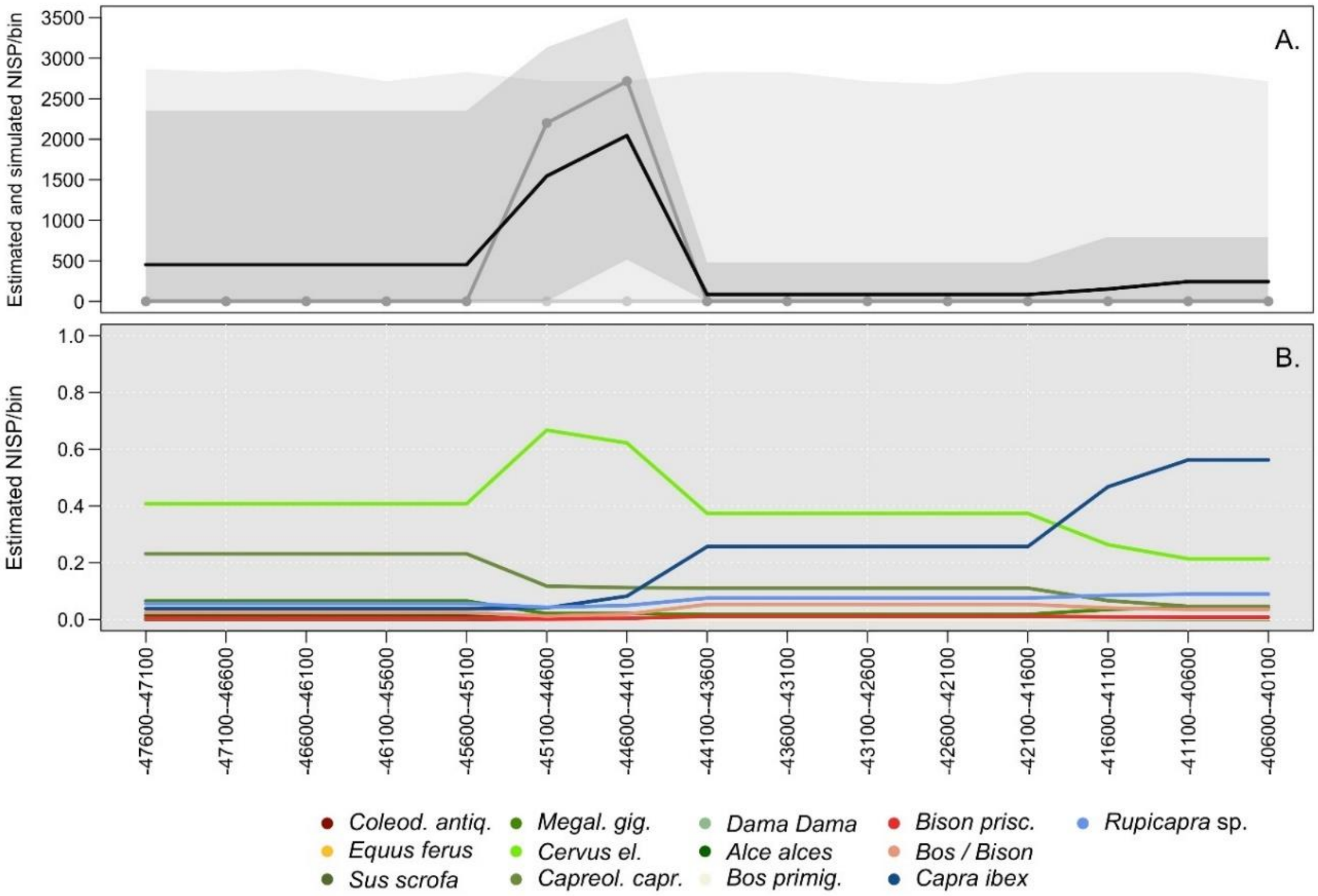

**Fig. 9.** A) Aoristic sum and estimated chronological frequency distribution of ungulates at Grotta di Fumane. The solid black line represents the aoristic sum, while the darker grey polygon indicates the simulated 95% confidence interval. The dark grey solid line-and-dots corresponds to simulated median values. The lighter grey envelope is instead the 95% confidence region of the null model based on uniform frequency distribution across the entire study period; B) Estimated diachronic relative frequency of each ungulate taxon based on aoristic sums computed for 500-year temporal bins. Horizontal axis indicates dates cal. BP.

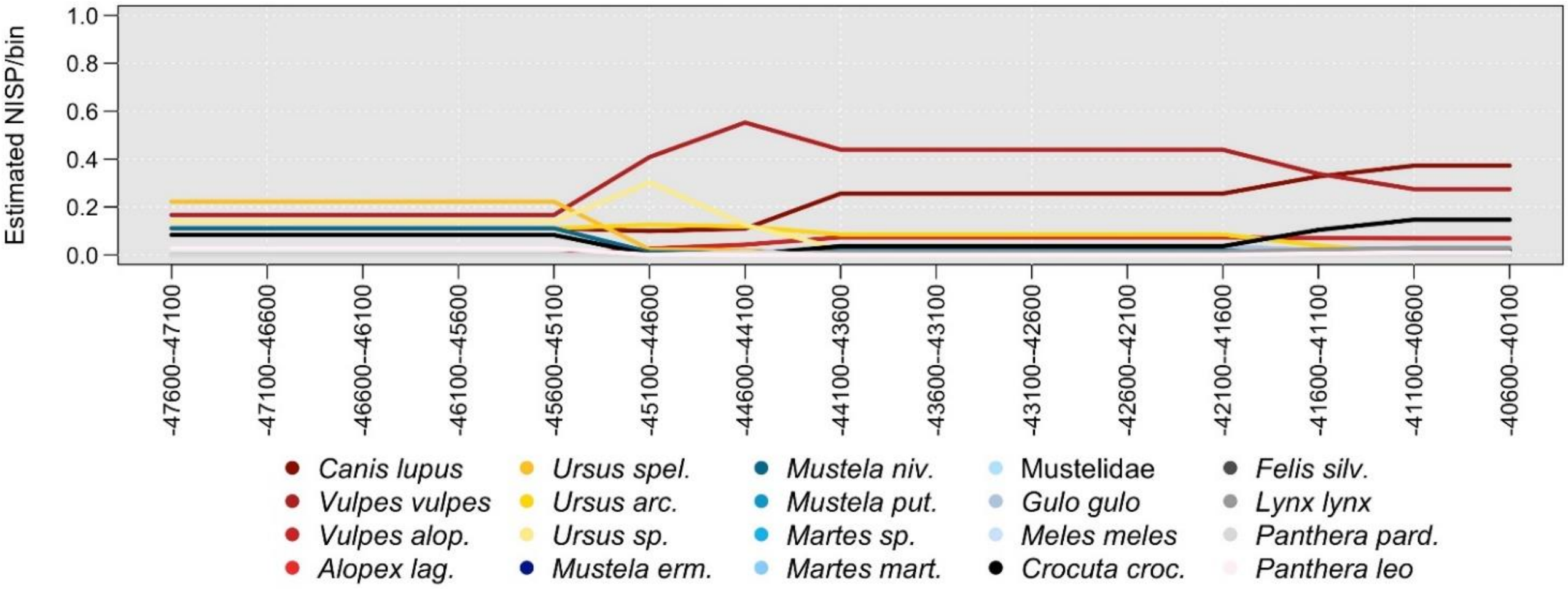

**Fig. 10.** Estimated diachronic relative frequency of each carnivore taxon at Grotta di Fumane, based on aoristic sums computed for 500-year temporal bins. Horizontal axis indicates dates cal. BP.

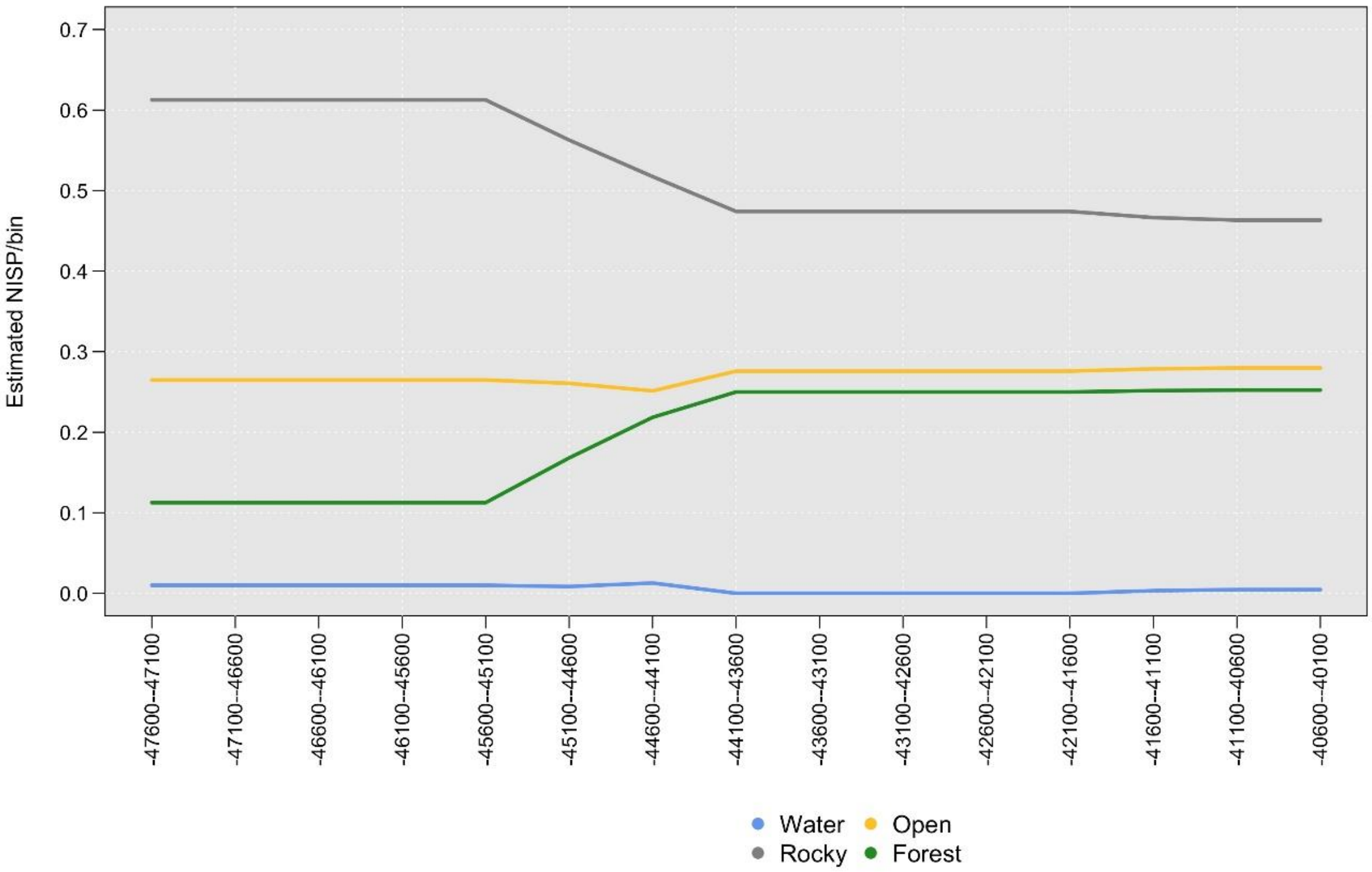

**Fig. 11.** Estimated diachronic relative frequency of avifaunal remains grouped by the relative environment, based on aoristic sums computed for 500-year temporal bins. Horizontal axis indicates dates cal. BP.

At the end of the Middle Palaeolithic (Fumane A6, A5+A6; San Bernardino, Unit II), Ursidae (*Ursus spelaeus* and *Ursus arctos*) were heavily exploited for fur, meat and marrow, while red fox and beaver were hunted for their skins (Fig. 13). In the Uluzzian at Fumane (A3) there is clear evidence of skinning of foxes, *Canis lupus*, and *Ursus arctos*. (Tagliacozzo et al., 2013; Romandini, 2012; Romandini et al., 2014a, 2016a, 2018a, b). At the same site, the Protoaurignacian (A2) shows evidence of anthropic exploitation of all these carnivores (Fig. 13) with the addition of Eurasian lynx (*Lynx lynx*), while until now there is no evidence of exploitation of avifauna outside of the Mousterian and Uluzzian deposits (Peresani et al., 2011a; Tagliacozzo et al., 2013; Romandini et al., 2014b, 2016b; Fiore et al., 2004, 2016).

The southern Italian assemblages show the same high proportion of ungulates bearing traces of human exploitation as they do in northern Italy (Fig. 13). In addition, also in southern Italian contexts, the spectrum of hunted species mirrors coeval changes in climate and environment, and anthropic modifications are aimed at obtaining skins, meat, and marrow. It is interesting to note the low number of butchered small carnivores and lagomorph taxa, which are particularly concentrated in the Uluzzian and Protoaurignacian phases (Fig. 13). Grotta della Cala in southwestern/Tyrrhenian Italy exhibits increasingly higher percentages of phalanges and sesamoids across the Uluzzian, Protoaurignacian, Early Gravettian and Evolved Gravettian (Benini et al., 1997; Boscato et al., 1997; Boscato and Crezzini, 2006, 2012).

Taphonomic analysis was carried out on a sample of identified ungulate remains from the Late Mousterian (NISP = 67), Uluzzian (NISP = 251), and Protoaurignacian (NISP = 38) layers at Castelcivita (Southwestern Italy; Table 6). Specimens mostly consist of cranial and limb bones (especially metacarpals and metatarsals), while evidence of vertebral bones is limited. The frequency of small limb bones (phalanges, sesamoids, carpal, and tarsal) is higher in Uluzzian and Protoaurignacian layers than in Mousterian ones (Mousterian = 17.9%, Uluzzian = 20.3%, Protoaurignacian = 25%). The ratio of diaphysis/epiphysis fragments is lower in the Uluzzian and Protoaurignacian (4) than in the Late Mousterian (5.3). Anthropic marks were identified on 7.5% of remains in the Late Mousterian sample, on 10.6% of Uluzzian material, and have not been identified in the Protoaurignacian assemblage. Carnivore gnawing marks are fewer in the Protoaurignacian layers (2.6% of total material) and more abundant in the Late Mousterian (4.5%) and Uluzzian (7.7%) layers.

A sample of unidentified remains from the Mousterian levels of Castelcivita (n = 1920) is highly fragmented (91.9% falls in the class 1–3 cm) (Table 5). In this context diaphysis fragments are the most abundant (40.4%), followed by spongy bones (16%), rib fragments (14.7%), and epiphysis fragments (7.2%) (Tables A.10 and A.11). Anthropic marks are present on 2.8% of the unidentified material, while carnivores left traces on 1.5% of the remains (mostly vertebrae and epiphysis fragments).

As far as skeletal components of the most represented taxa are concerned (Table 7), results obtained at Castelcivita are consistent with what emerged from other southern contexts (Boscato and Crezzini, 2006, 2012), i.e. small limb bones (phalanges, sesamoids, carpals, and tarsals) and epiphyses are present with increasingly higher frequency across the archaeological sequence, while diaphysis fragments exhibit increasingly lower frequency (Tables 7, A.10 and A.11). Nevertheless, the estimate of the contribution of anthropic actions to the formation of faunal assemblages found at Castelcivita may be biased by the presence of spotted hyena (*Crocuta crocuta spelaea*).

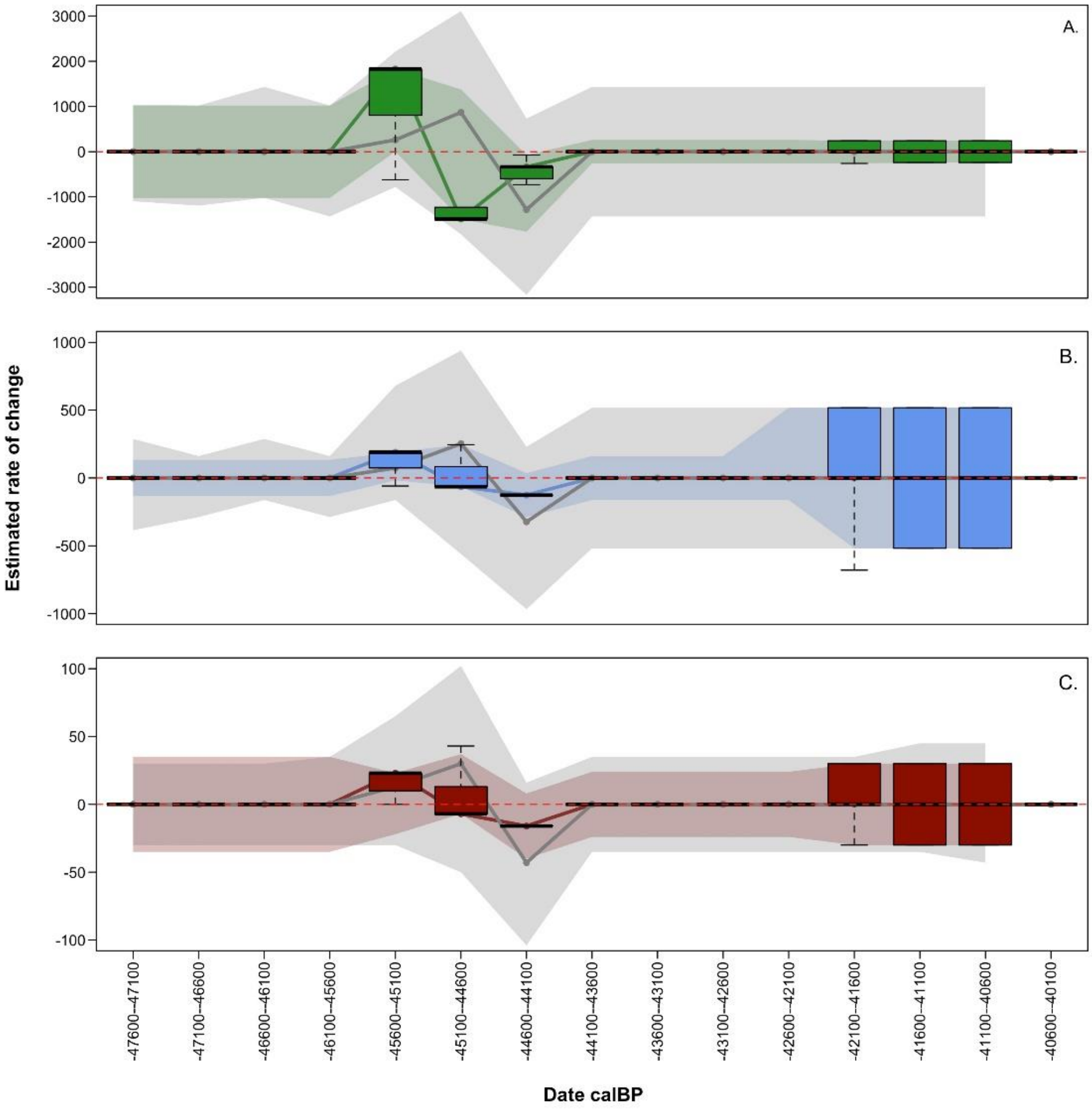

**Fig. 12.** Simulated diachronic rates of change computed for ungulate families at Grotta di Fumane. Box-plots and coloured polygons indicate the distribution of simulated rates of change based on observed family frequencies, while grey polygons represent 95% confidence regions for a null model based on the aoristic sums of carnivores uncovered at the same archaeological site. A (green): Cervidae; B (blue): Caprinae; C (red): Bovinae. Based on 5000 simulation runs and computed at 500-year bin resolution. (For interpretation of the references to color in this figure legend, the reader is referred to the Web version of this article.)

Most unidentified specimens fall in the smallest dimensional category (1–3 cm), while the percentage of larger findings is higher in Uluzzian deposits (Table 5). Turning to Southeastern Italy, fragments of long bone diaphyses are abundant in the Mousterian assemblages of Riparo l'Oscurusciuto and Grotta del Cavallo whereas epiphysis fragments are rare. At Grotta del Cavallo, on the other hand, percentages of diaphysis fragments are lower in the Uluzzian than they are in the Late Mousterian ones (Table A.11; Boscato and Crezzini, 2006, 2012). Considering Riparo Oscurusciuto and Grotta del Cavallo (where bone assemblages are not biased by the action of carnivores), the observed differences between the Late Mousterian and the Uluzzian in the proportion of diaphysis, spongy-bone and phalanges are statistically significant (Tables A.14, A.15, A.16 and A.17). As far as the degree of fragmentation is concerned, it is important not to directly compare any of the (preliminary) values currently available for Southern assemblages with those presented for the northern regions.

## 5. Discussion

### *5.1. Comparison of taxon frequencies in macro-mammals between Northeastern, Southwestern, and Southeastern Italy*

Mammal assemblages show that the Middle to Upper Palaeolithic Transition in Northern Italy was associated with a shift to colder and arid climatic conditions, as previously observed by Fiore et al. (2004) and Holt et al. (2019). In Northeastern Italy, human groups used rock shelters in the prealpine fringe and in the alpine foreland and exploited closed forest environments. The surroundings of such shelters were characterised by open environments, alpine meadows and cliffs populated by herbaceous and shrubby species, while humans had to share and compete for their shelters with bears (Romandini et al., 2018a). At the end of Middle Palaeolithic, the examined faunal assemblages are dominated by cervidae while species adapted to open environments became considerably less abundant, suggesting a gradual change towards more temperate-humid climate which favoured the expansion of forests and wooded environments before the Uluzzian (such as in Fumane A4). Uluzzian and Protoaurignacian (e.g. Fumane A2) layers bear instead evidence of an abrupt shift to colder and arid conditions, which favoured the diffusion of steppic environments and alpine meadows. These observations are supported by the relative frequencies of cervids and caprids, both of which appear in higher percentages in moments of higher absolute intensity of deposition of ungulate remains.

Caprids and bovids also show instances of increase and decrease that are not entirely predicted by the null model based on the distribution of carnivores (i.e. might actually be related to change in environmental conditions).

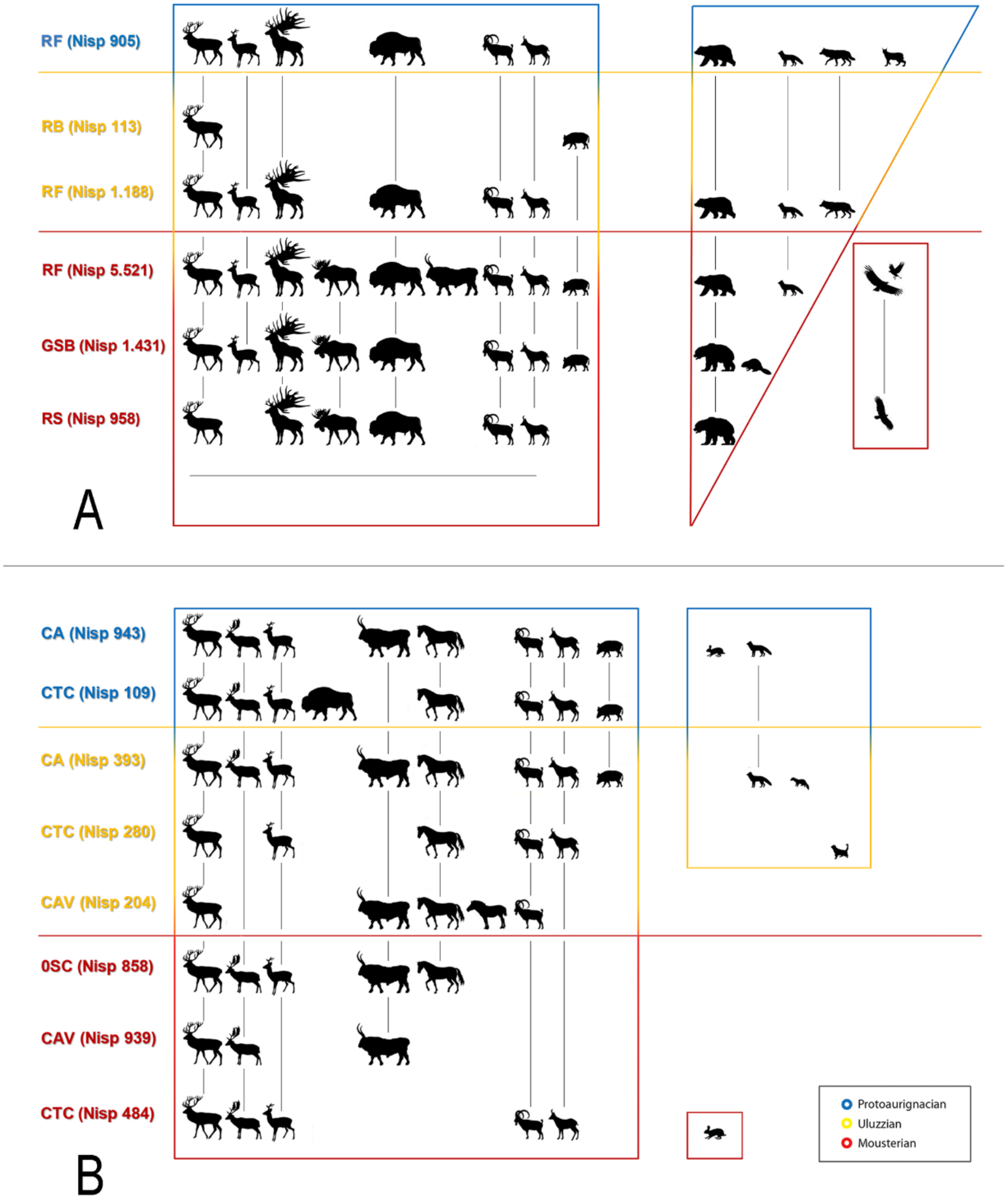

**Fig. 13.** Comparative summary of diachronic trends concerning the presence of taxa whose bones bear cut-marks, percussion marks and/or evidence of combustion. A) Northern Italy. RF = Grotta di Fumane; RB = Riparo del Broion; GSB = Grotta di San Bernardino; RS = Grotta del Rio Secco. B) Southern Italy. CAV = Grotta del Cavallo; OSC = Riparo l'Oscurusciuto; CTC = Grotta di Castelcivita; CA = Grotta della Cala. Colors of temporal phases: Protoaurignacian = blue; Uluzzian = yellow; Late Mousterian = red (for interpreting the color coding in the figure legend the reader is referred to the Web version of this article).

The aoristic analysis of Grotta di Fumane's zooarchaeological data confirm some of the trends observed by investigating taxon frequency across different sites of Northeastern Italy, especially in the first half of the study sequence. In addition, the comparison of simulated trends against null models based on constant deposition and on the aoristic sum of carnivores provides a means to more formally assess empirical patterns against explicit scenarios. This is particularly useful in a case study affected by small sample size and limited data comparability such as the present one. Most trends appear flattened in the aoristic sum graph, since the analysis explicitly incorporates the temporal uncertainty embedded in the present dataset. Nevertheless, the adoption of this approach paves the ground for future direct comparisons between northern and southern contexts by highlighting long-term processes that can be directly compared against palaeoecological and palaeoclimatic data collected from a variety of archives, so that future inferences on change in adaptive strategies can be more objective. Additional dates and more detailed information on taphonomy and post-depositional processes will also help shed light on the mechanisms actually underlying the potential diachronic change for Protoaurignacian contexts.

| MODIF. | RS 5+8 LM | | RF A9 LM | | RF A6 LM | | RF A5 LM | | RS 5top+7 LM | | SB II+III LM | | RF A4 LM | | RB 1e+1f+1g UL | | RF A3 UL | | RF A2 PA | |
|---|---|---|---|---|---|---|---|---|---|---|---|---|---|---|---|---|---|---|---|---|
| | NR | % | NR | % | NR | % | NR | % | NR | % | NR | % | NR | % | NR | % | NR | % | NR | % |
| CM+SCR | 61 | 63.5 | 882 | 53.5 | 1003 | 37.8 | 399 | 35.5 | 76 | 67.3 | 92 | 54.8 | 626 | 68.9 | 16 | 59.3 | 289 | 53.4 | 348 | 64.7 |
| CM+IF - CM+PM | 10 | 10.4 | 143 | 8.7 | 171 | 6.5 | 50 | 4.4 | 20 | 17.7 | 14 | 8.3 | 128 | 14.1 | | | 106 | 19.6 | 30 | 5.6 |
| IF+PM | 25 | 26 | 623 | 37.8 | 1477 | 55.7 | 676 | 60.1 | 17 | 15 | 62 | 36.9 | 154 | 17 | 11 | 40.7 | 146 | 27 | 160 | 29.7 |
| TOTAL BM | 96 | | 1648 | | 2651 | | 1125 | | 113 | | 168 | | 908 | | 27 | | 541 | | 538 | |

**Tab. 3** Number of remains (NR) and relative % with anthropic modifications identified in the MP/UP transition from Northern Italian sites (see Fig.1 - Area 1). RF = Grotta di Fumane; RB = Riparo del Broion; SB = Grotta di San Bernardino; RS = Grotta del Rio Secco. CM = Cut Marks; SCR. = Scrapings; IF= Impact Flakes; PM= Percussion Marks; BM = Butchering Marks.

In southern Italy, Late Mousterian deposits exhibit evidence of generally temperate conditions. In this phase, cervids are the most common ungulates in the Tyrrhenian region, while *Bos primigenius* is the most represented species in Ionian contexts. At Grotta di Castelcivita, this phase is characterised by the same palaeoenvironmental trend documented at Fumane.

The Uluzzian phases at Grotta del Cavallo and Grotta di Castelcivita show data compatible with the establishment of a colder climate, while human groups active at Grotta della Cala experienced more temperate conditions. During MIS 3, the Tyrrhenian side appears to be still characterised by temperate and humid conditions favouring forests and wooded environments, while the Ionian areas were marked by steppic environments and wooded steppe.

Outside of the Italian Peninsula, the only possible comparisons in terms of temporal span and of an archaeological sequence comprising Late Mousterian – Uluzzian – Protoaurignacian (only Aurignacian in Greece) It's represented by Kephalari and Klissoura Cave, Peloponnese, Greece (Starkovich, 2012; Starkovich and Ntinou, 2017; Starkovich et al., 2018). Especially at the latter site, the MIS 3 was highly variable, as suggested by evidence of variation between forested environments, mixed forest-steppe (with red deer, roe deer, chamois and ibex), and drier intervals with steppe species (such as European wild ass, aurochs, ibex and great bustard). The Uluzzian (V) and Aurignacian (IV) layers yielded evidence of fallow deer and small game, in addition to species adapted to both open and forested environments. Plants indicate a mixture of forest and steppe, although taxonomic evenness suggests that conditions were slightly wetter in the Uluzzian layers than during the final MP occupations.

If we exclude Upper Palaeolithic layers of both Kephalari and Klissoura Cave, the exploitation of small game across the transition between Middle and Upper Palaeolithic could be linked to coeval environmental change and a change in resource availability, as it is suggested by the remains of tortoise and hare identified at Klissoura Cave1 in assemblages associated with Neandertals (Starkovich, 2012, 2017, Starkovich et al., 2018). The range of hunted taxa in this region therefore seems to be stable across the Middle and Upper Palaeolithic, and trends can be ascribed to species availability dictated by environmental and climatic change, rather than to convergence in hunting strategies with the Italian Peninsula (Starkovich et al., 2018; Stiner and Munro, 2011). At present a more detailed comparison between the exploitation of animal remains documented in Italy and Greece is not yet possible as research on the subject is still in progress and additional data are required. Nevertheless, trends emerging from taxon abundance analysis are broadly comparable to those identified for Southwestern and Southeastern Italy and documented in the present work. On the other hand, preliminary results presented here suggest in Middle to UP transition a more intensive exploitation of small game in Northeastern Italy than in Southern contexts and the Peloponnese. This finding might be particularly relevant for interpreting regional patterns of change in subsistence/adaptive strategies, considering that large game is generally considered a higher-rank resource than small game.

### *5.2. Comparison of avifaunal remains between Fumane and Castelcivita*

The avifaunal assemblages of Grotta di Fumane and Grotta di Castelcivita provide relevant insights on the paleoenvironmental and palaeoclimatic framework of both deposits. The surroundings of Fumane were characterised by mixed and conifer forests, grasslands and alpine meadows with rocky outcrops, cliffs and slow-flowing water bodies. The environmental framework of Grotta di Castelcivita was instead characterised by wetlands in the plains in front of the cave and by drier habitats like grasslands, rocky terrains and rock walls, alternated to conifer or mixed forests at higher altitudes.

Bird taxa across Italy indicate the presence of a consistently colder climate than the present one. Nevertheless, in the southwestern/ Tyrrhenian area (Fig. 1, Area 2) climate seems milder and more temperate than in the Adriatic area, as suggested by the absence of boreal bird species and by a lower proportion of high altitude bird taxa in the former. Once again, faunal assemblages mirror climatic differences between Tyrrhenian and Ionian/Adriatic regions mostly due to the effect of the Balkanic influence on the latter. The Middle Paleolithic assemblages from both sites provide evidence of temperate-cool climate, where the species of open and rocky habitats prevail. The Late Mousterian Layer A6 at Fumane hints at a possible cold oscillation, however, and the Uluzzian at both sites (A3 at Grotta di Fumane, and CTC rsi at Grotta di Castelcivita) shows higher percentages of bird taxa typical of open habitats possibly due to colder conditions linked to Heinrich Event 4 (Higham et al., 2009; Moroni et al., 2018; Lopéz-García et al., 2015). Protoaurignacian deposits provide evidence for the persistence of harsh conditions which characterized previous phases (Cassoli and Tagliacozzo, 1994a). In spite of the low NISP, one exception seems to be represented by the latest Protoaurignacian layers of Castelcivita (gic-ars; referring here particularly to the layer ars, as the bird bone sample of the layer gic was too small to include it in the analysis), that yielded evidence for climatic shift toward more humid conditions (Cassoli and Tagliacozzo, 1997; Gala et al., 2018).

| TECHNOCOMPLEX | LM | | LM | | LM | | LM | | LM | | LM | | LM | | LM | | LM | | LM | | LM | | LM | | UL | | UL | | UL | | PA | |
|---|---|---|---|---|---|---|---|---|---|---|---|---|---|---|---|---|---|---|---|---|---|---|---|---|---|---|---|---|---|---|---|---|
| SITE - US/Lev. | RS 5+8 | | RF A9 | | RF A9 | | RF A6 | | RF A6 | | RF A5 | | RF A5 | | RS 5top+7 | | SB II+III | | SB II+III | | RF A4 | | RF A4 | | RB 1f+1g | | RF A3 | | RF A3 | | RF A2 | |
| *Taxa* | *Ursus* sp. | | *C. elaphus* | | *C. capreolus* | | *C. elaphus* | | *C. capreolus* | | *C. elaphus* | | *C. capreolus* | | *Ursus* sp. | | *C. elaphus* | | *C. capreolus* | | *C. elaphus* | | *Capra ibex* | | *Sus scrofa* | | *C. elaphus* | | *Capra ibex* | | *Capra ibex* | |
| | Nisp | % | Nisp | % | Nisp | % | Nisp | % | Nisp | % | Nisp | % | Nisp | % | Nisp | % | Nisp | % | Nisp | % | Nisp | % | Nisp | % | Nisp | % | Nisp | % | Nisp | % | Nisp | % |
| Cranium | 14 | 6.4 | 23 | 4.6 | 7 | 2.5 | 43 | 3.8 | 6 | 3.3 | 10 | 3.2 | 1 | 2 | 18 | 8.3 | 15 | 11.3 | 18 | 7.5 | 10 | 4.6 | 7 | 9.5 | 1 | 4.2 | 7 | 3.8 | 15 | 12.9 | 6 | 4.6 |
| Emimandible | 21 | 9.6 | 35 | 7.1 | 16 | 5.7 | 64 | 5.6 | 6 | 3.3 | 19 | 6.1 | 7 | 14 | 13 | 6 | 32 | 24.1 | 36 | 15.1 | 14 | 6.4 | 15 | 20.3 | 1 | 4.2 | 10 | 5.5 | 22 | 19 | 13 | 10 |
| Tooth indet. | 14 | 6.4 | 14 | 2.8 | 6 | 2.1 | 17 | 1.5 | 1 | 0.5 | 1 | 0.3 | | | 2 | 0.9 | | | 8 | 3.3 | 13 | 6 | 3 | 4.1 | 3 | 12.5 | 24 | 13.2 | 3 | 2.6 | 7 | 5.4 |
| Hioyd | 6 | 2.8 | | | | | 1 | | | | 2 | 0.6 | | | 2 | 0.9 | 1 | 0.8 | | | | | | | | | 2 | 1.1 | | | 1 | 0.8 |
| Atlas-axis | 2 | 0.9 | | | | | 1 | 0.1 | | | | | | | 3 | 1.4 | | | | | | | | | | | | | | | | |
| Vertebra | 21 | 9.6 | 3 | 0.6 | 2 | 0.7 | 7 | 0.6 | 1 | 0.5 | 2 | 0.6 | | | 12 | 5.6 | | | | | 2 | 0.9 | | | | | 2 | 1.1 | 1 | 0.9 | 3 | 2.3 |
| Rib | 39 | 17.9 | 4 | 0.8 | 2 | 0.7 | 7 | 0.6 | | | 1 | 0.3 | | | 54 | 25 | 1 | 0.8 | | | 1 | 0.5 | | | | | | | | | | |
| Clavicle | | | | | | | | | | | | | | | 1 | 0.5 | | | | | | | | | | | | | | | | |
| Baculum | 1 | 1.6 | | | | | | | | | | | | | 2 | 0.9 | | | | | | | | | | | | | | | | |
| Scapula | 3 | 6.1 | | | | | 4 | 0.4 | 1 | 0.5 | | | | | 3 | 1.4 | | | | | | | 1 | 1.4 | | | 3 | 1.6 | | 0 | 2 | 1.5 |
| Humerus | 2 | 0.9 | 27 | 5.5 | 12 | 4.3 | 89 | 7.8 | 5 | 2.7 | 24 | 7.7 | 1 | 6.25 | 3 | 1.4 | 5 | 3.8 | 2 | 0.8 | 6 | 2.8 | 4 | 5.4 | | | 9 | 4.9 | 4 | 3.4 | 6 | 4.6 |
| Radius/Ulna | | | | | | | 3 | | | | 1 | 0.3 | | | | | 3 | 2.3 | | | | | 2 | 2.7 | | | | | 1 | 0.9 | | |
| Radius | 3 | 1.4 | 29 | 5.9 | 10 | 3.6 | 69 | 6.1 | 6 | 3.3 | 14 | 4.5 | 1 | 2 | 15 | 6.9 | 3 | 2.3 | 2 | 0.8 | 10 | 4.6 | 2 | 2.7 | 1 | 4.2 | 5 | 2.7 | 3 | 2.6 | 6 | 4.6 |
| Ulna | 6 | 2.8 | 4 | 0.8 | 1 | 0.4 | 26 | 2.3 | 3 | 1.6 | 3 | 1 | | | 4 | 1.9 | 4 | 3 | 2 | 0.8 | 6 | 2.8 | | | | | 2 | 1.1 | 1 | 0.9 | 2 | 1.5 |
| Carpals | 6 | 2.8 | | | 6 | 2.1 | 8 | 0.7 | 5 | 2.7 | 1 | 0.3 | | | 7 | 3.2 | 1 | 0.8 | 1 | 0.4 | 1 | 0.5 | 4 | 5.4 | | | 1 | 0.5 | 9 | 7.8 | 3 | 2.3 |
| Metacarpal | 4 | 1.8 | 55 | 11.1 | 32 | 11.4 | 123 | 10.8 | 19 | 10.4 | 40 | 12.9 | 11 | 22 | 12 | 5.6 | 6 | 4.5 | 12 | 5 | 19 | 8.7 | 2 | 2.7 | 1 | 4.2 | 16 | 8.8 | 6 | 5.2 | 4 | 3.1 |
| Metacarpal rud. | | | | | 4 | 1.4 | 6 | 0.5 | 1 | 0.5 | | | 1 | 2 | | | | | 5 | 2.1 | | | | | | | | | | | 4 | 3.1 |
| Coxal | | | 4 | 0.8 | | | 5 | 0.4 | 2 | 1.1 | 1 | 0.3 | | | | | 1 | 0.8 | | | 1 | 0.5 | 2 | 2.7 | | | 2 | 1.1 | 3 | 2.6 | | |
| Femur | 28 | 12.8 | 44 | 8.9 | 17 | 6 | 117 | 10.3 | 11 | 6 | 27 | 8.7 | 3 | 6 | 24 | 11.1 | 6 | 4.5 | 3 | 1.3 | 17 | 7.8 | 2 | 2.7 | | | 9 | 4.9 | 10 | 8.6 | 5 | 3.8 |
| Patella | 1 | 0.5 | | | 1 | 0.4 | | | | | 1 | 0.3 | 1 | 2 | | | | | 1 | 0.4 | | | | | | | | | | | | |
| Tibia | 10 | 4.6 | 96 | 19.4 | 38 | 13.5 | 181 | 15.9 | 15 | 8.2 | 49 | 15.8 | | | 8 | 3.7 | 8 | 6 | 3 | 1.3 | 16 | 7.3 | 4 | 5.4 | | | 18 | 9.9 | 5 | 4.3 | 6 | 4.6 |
| Fibula | 4 | 1.8 | | | | | | | | | | | | | 4 | 1.9 | | | | | | | | | | | | | | | | |
| Malleolar bone | | | | | | | 2 | 0.2 | 2 | 1.1 | | | | | | | | | | | 2 | 0.9 | 1 | 1.4 | | | | | | | 1 | 0.8 |
| Calcaneum | | | 1 | 0.2 | | | | | 1 | 0.5 | | | | | 2 | 0.9 | | | 1 | 0.4 | 1 | 0.5 | | | | | | | | | 1 | 0.8 |
| Astragalus | | | | | | | 1 | 0.1 | 2 | 1.1 | | | | | 1 | 0.5 | 1 | 0.8 | 1 | 0.4 | | | | | | | 1 | 0.5 | | | 3 | 2,3 |
| Tarsals | 1 | 0.5 | 2 | 0.4 | 1 | 0.4 | 3 | 0.3 | 2 | 1.1 | 2 | 0.6 | | | 3 | 1.4 | 2 | 1.5 | 1 | 0.4 | | | 2 | 2.7 | | | | | 1 | 0.9 | 4 | 3.1 |
| Metatarsal | 5 | 2.3 | 76 | 15.4 | 59 | 21 | 131 | 11.5 | 29 | 15.9 | 54 | 17.4 | 8 | 16 | 1 | 0.5 | 12 | 9 | 28 | 11.7 | 41 | 18.8 | 2 | 2.7 | 3 | 12.5 | 29 | 15.9 | 4 | 3.4 | 3 | 2.3 |
| Metapodial | 2 | 0.9 | 25 | 5.1 | 11 | 3.9 | 45 | 3.9 | 6 | 3.3 | 13 | 4.2 | 1 | 2 | | | 2 | 1.5 | 9 | 3.8 | 9 | 4.1 | | | 3 | | 10 | 5.5 | 4 | 3.4 | 2 | 1.5 |
| First phal. | 11 | 5 | 11 | 2.2 | 21 | 7.5 | 28 | 2.5 | 7 | 3.8 | 10 | 3.2 | 1 | 2 | 8 | 3.7 | 8 | 6 | 35 | 14.6 | 7 | 3.2 | 8 | 10.8 | 3 | 12.5 | 4 | 2.2 | 4 | 3.4 | 12 | 9.2 |
| Second phal. | 6 | 2.8 | 12 | 2.4 | 9 | 3.2 | 50 | 4.4 | 15 | 8.2 | 16 | 5.1 | 6 | 12 | 8 | 3.7 | 9 | 6.8 | 24 | 10 | 15 | 6.9 | 3 | 4.1 | 3 | 12.5 | 10 | 5.5 | 3 | 2.6 | 9 | 6.9 |
| Third phal. | 6 | 2.8 | 4 | 0.8 | 4 | 1.4 | 28 | 2.5 | 1 | 0.5 | 6 | 1.9 | 2 | 4 | 6 | 2.8 | 3 | 2.3 | 13 | 5.4 | 4 | 1.8 | 1 | 1.4 | 1 | 4.2 | 5 | 2.7 | 2 | 1.7 | 6 | 4.6 |
| First phal. rud. | | | 3 | 0.6 | 4 | 1.4 | 9 | 0.8 | 4 | 2.2 | | | | | | | 1 | 0.8 | 4 | 1.7 | 2 | 0.9 | | | 3 | 12.5 | 2 | 1.1 | | | | |
| Sec. phal. rud. | | | 5 | 1 | 4 | 1.4 | 14 | 1.2 | 7 | 3.8 | 3 | 1 | 2 | 4 | | | | | 7 | 2.9 | 8 | 3.7 | | | | | 4 | 2.2 | | | | |
| Th. phal. rud. | | | 4 | 0.8 | 7 | 2.5 | 9 | 0.8 | 3 | 1.6 | 3 | 1 | | | | | 3 | 2.3 | 4 | 1.7 | 2 | 0.9 | | | | | 3 | 1.6 | | | | |
| Sesamoid | 2 | 0.9 | 14 | 2.8 | 7 | 2.5 | 49 | 4.3 | 21 | 11.5 | 8 | 2.6 | 4 | 8 | | | 6 | 4.5 | 19 | 7.9 | 11 | 5 | 9 | 12.2 | 1 | 4.2 | 4 | 2.2 | 15 | 12.9 | 21 | 16.2 |
| TOTAL | 218 | | 495 | | 281 | | 1140 | | 182 | | 311 | | 50 | | 216 | | 133 | | 239 | | 218 | | 74 | | 24 | | 182 | | 116 | | 130 | |
| Tot Cranium + tooth | *55* | *25.2* | *72* | *14.5* | *29* | *10.3* | *125* | *11* | *13* | *7.1* | *32* | *10.3* | *8* | *16* | *35* | *16.2* | *48* | *36.1* | *62* | *25.9* | *37* | *17* | *25* | *33.8* | *5* | *20.8* | *43* | *23.6* | *40* | *34.5* | *27* | *20.8* |
| Tot trunk | *63* | *28.9* | *7* | *1.4* | *4* | *1.4* | *15* | *1.3* | *1* | *0.5* | *3* | *1* | | | *72* | *33.3* | *1* | *0.8* | | | *3* | *1.4* | | | | | *2* | *1.1* | *1* | *0.9* | *3* | *2.3* |
| Tot long limb bones | *68* | *35.9* | *360* | *72.9* | *185* | *65.8* | *799* | *70.1* | *98* | *56.6* | *227* | *73* | *27* | *58.3* | *74* | *35.6* | *50* | *38.3* | *67* | *28.9* | *125* | *58.7* | *21* | *29.7* | *8* | *20.8* | *103* | *57.1* | *41* | *35.3* | *40* | *34.6* |
| Carpal + tarsal | *7* | *3.2* | *3* | *0.4* | *7* | *2.5* | *14* | *1* | *12* | *3.8* | *3* | *1* | | | *13* | *4.6* | *4* | *2.3* | *4* | *0.8* | *4* | *0.5* | *7* | *8.1* | | | *2* | *0.5* | *10* | *8.6* | *12* | *5.4* |
| Phalan + sesamoides | *25* | *11.5* | *53* | *10.7* | *56* | *19.9* | *187* | *16.4* | *58* | *31.9* | *46* | *14.8* | *15* | *30* | *22* | *10.2* | *30* | *22.6* | *106* | *44.4* | *49* | *22.5* | *21* | *28.4* | *11* | *45.8* | *32* | *17.6* | *24* | *20.7* | *48* | *36.9* |

**Tab. 4** Number of remains and relative % of the specific anatomical elements, referring to the most represented mammals species present in the layers and levels analysed in MP/UP transition Northern Italian sites (see Fig. 1 – Area 1). The subtotals of the different anatomical compartments are reported at the bottom of the table. RF = Grotta di Fumane; RB = Riparo del Broion; SB = Grotta di San Bernardino; RS = Grotta del Rio Secco.

### *5.3. Taphonomy*

Despite the facts that taphonomic data are still only partially investigated in most of the sampled contexts and that the majority of identified patterns cannot be proven to be statistically significant because of small sample size, interesting preliminary trends emerge. Although future studies may alter the pattern detected so far, at present, the percentage of calcined remains during the Uluzzian and Protoaurignacian levels in Northeastern Italian contexts is higher than the frequency of the same items in previous phases, and hints at a possible behavioural change linked to the use of fire: greater intensity and duration of use of the hearths, differentiation of fuel and/or cooking of animal resources. Cut-marks are also more frequent across the transition, while the degree of bone fragmentation for marrow extraction is higher in Mousterian layers than in later deposits. In the Early Upper Palaeolithic overall (i.e., Uluzzian and Protoaurignacian) there are higher percentages of cranial bones and limb extremities, with a consequent lower proportion of long bones. This trend may be imputed partly to human selection and partly to the use of the cave by hyenas and other carnivores. The remains of the most frequently hunted large (Cervidae, Bovinae) and medium-sized (Caprinae) ungulates show cut- and percussion-marks, all of which point to skinning, butchering, and marrow extraction. Over the same timespan, bears and middle- and small-sized carnivores appear to be more frequently exploited, suggesting a broadening in the range of species hunted for skin and fur (Collard et al., 2016).

| SITE<br>US - Levels | Technocomplex | 1 - 3 cm | % | > 3 cm | % | TOTAL Rem. | Burn.+Calc. | % |
|---|---|---|---|---|---|---|---|---|
| CAV EIII | UL | 4201 | 79.9 | 984 | 20.1 | 5185 | 3452 | 82.2 |
| CTC LM | LM | 1764 | 91.9 | 156 | 8.1 | 1920 | Not avail. | Not avail. |
| CAV FII | LM | 9836 | 87.7 | 1378 | 12.3 | 11214 | 1744 | 17.7 |
| OSC US 4/1 | LM | 17472 | 97.4 | 449 | 2.6 | 17921 | 12137 | 67.7 |

**Tab. 5** Size classes of mammals bones and relative frequency of burnt remains identified in the LM and UL layers sampled in Southern Italian sites (see Fig. 1 – Area 2 + 3). CAV = Grotta del Cavallo; CTC = Grotta di Castelcivita; OSC = Grotta l'Oscurusciuto.

| | CTC LM | | CAV FII LM | | OSC US 4 LM | | CAV EII5 UL | | CTC UL | | CTC PA | |
|---|---|---|---|---|---|---|---|---|---|---|---|---|
| MODIF. | NR | % | NR | % | NR | % | NR | % | NR | % | NR | % |
| CM+SCR | 9 | 0.4 | 63 | 67 | 1 | 3.3 | 75 | 80.6 | 16 | 89 | 5 | 17.2 |
| CM+IF / CM+PM | 43 | 45.3 | 13 | 13.8 | 26 | 83.9 | 11 | 11.8 | 1 | 5.5 | 12 | 41.4 |
| IF+PM | 43 | 45.3 | 18 | 19.2 | 4 | 2.8 | 7 | 0.6 | 1 | 5.5 | 12 | 41.4 |
| TOTAL BM | 95 | | 94 | | 31 | | 93 | | 18 | | 29 | |

**Tab. 6** Number (NR) and relative frequency (%) of remains with identified anthropic modification documented in the MP/UP transitional contexts from the sampled Southern Italian sites (see Fig. 1 – Area 2 + 3). CAV = Grotta del Cavallo; OSC = Riparo l'Oscurusciuto; CTC = Grotta di Castelcivita. CM = Cut Marks; SCR. = Scrapings; IF= Impact Flakes; PM= Percussion Marks; BM = Butchering Marks.

Avifaunal assemblages provide evidence of human consumption of birds and contribute to an understanding of the role of avifaunal resources in the subsistence strategies of Middle Palaeolithic hominins (Peresani et al., 2011a; Romandini, 2012; Tagliacozzo et al., 2013; Fiore et al., 2016; Romandini et al., 2014b, 2016a, b; Gala et al., 2018; Fiore et al., 2004, 2016). The exploitation of these resources is testified by recognizable taphonomic indicators such as evidence for the exploitation of feathers from various raptors and other birds.

In the same way, evidence of Neandertal reliance on small mammal prey increased over the past 10 years due to the reassessment of faunal assemblages from a new taphonomic perspective (Romandini et al., 2018b; Morin et al., 2019).

In Ionian contexts, Late Mousterian assemblages exhibit a lack (or at least a scarcity) of long-bone epiphyses, carpal and tarsal bones, phalanges and sesamoides. In the analysed Late Mousterian samples from Grotta del Cavallo and Riparo l'Oscurusciuto, this evidence cannot be attributed to carnivores, differential bone density and other post-depositional processes (Boscato and Crezzini, 2006, 2012). The frequency of different anatomical parts (% of Minimum Animal Units, Binford, 1984) of *Bos primigenius* and the modal species in US 4 at Riparo l'Oscurusciuto were compared against Emerson's utility indices related to present-day *Bison bison* (Emerson, 1990, 1993). Previous results suggest a relationship between bone frequency and their content in marrow and fat, which was probably crucial in the choice to select specific anatomical parts and to carry them back to camps/sites (Boscato and Crezzini, 2006, 2012). Recent studies demonstrated that at least at Riparo l'Oscurusciuto spongy bones were not systematically used as fuel in hearths (Spagnolo et al., 2016), suggesting their possible use as food (Costamagno and Rigaud, 2014). In southern Italy, Upper Palaeolithic assemblages indicate a different manner of exploitating ungulate bones (Boscato and Crezzini, 2006, 2012). A large amount of long-bone epiphyses and spongy elements (including carpal and tarsal bones) were not destroyed and can be found in these assemblages. Unlike Neandertals, who were evidently not interested in phalanges and probably left them at the kill sites, modern humans usually transported these small skeletal parts to their campsites where they fragmented them to recover the particular fat they contained: Morin (2006) underlines that, although the phalanges contain a low quantity of marrow, it is qualitatively different than the marrow contained in long bones, due to its higher percentage of oleic acid. These data suggest a change in processing hard animal tissues by Upper Palaeolithic people across southern Italy, a change that is already visible in Uluzzian assemblages, a documented by the case of Layer EIII5 at Grotta del Cavallo (Boscato and Crezzini, 2006, 2012).

## 6. Conclusions

The data collected and analysed show that human adaptive strategies changed over time to cope with variability in local topographic and ecological conditions, as well as with uncertainty in resource availability. Uncertainty and bias are critically embedded in the procurement and exploitation of animal resources, especially in such a fragmented and small-scale scenario as the Italian Peninsula. A sample of key sites from both southern and northern Italy offers evidence of how groups of Neandertals and modern humans occupied the Tyrrhenian and Ionian areas, as well as through the area between the great alluvial plain of the river Po and the Pre-Alpine mountains. In this context, a thorough and detailed zooarchaeological approach offers a unique perspective on palaeoenvironmental and palaeoecological settings, as well as on hunting and subsistence strategies. In the present study, we presented state-of-the-art evidence on the differential presence of large mammals and avifauna across Late Mousterian, Uluzzian, and Protoaurignacian assemblages from Italy. Incorporating an aoristic analysis further allows us to explicitly address the amount of temporal uncertainty embedded in one of the zooarchaeological assemblages of interest. While losing detail on individual archaeological layers, this method offers a practical solution to help overcome the effects of time-averaging and of the lack of information on layer-specific accumulation rates. At the same time, aoristic sums for ungulates, rates of change based on simulated data, and the comparison with null models depict a conservative scenario useful for inferring instances of *absolute* increase or decline of given taxa or families over time. The analysis of environmentally-informative bird taxa added significant detail to the environmental trends provided by mammal remains, improving our understanding of the climatic framework of the Middle-Upper Paleolithic transition. The future addition of micromammals to the analysis will make it possible to add detail on local biotopes, and to further test inferences on palaeoclimatic change in the different contexts. Interesting hypotheses on human behavioural ecology also emerge from the examined archaeological assemblages, although additional evidence is still clearly required for objectively test inferences about Uluzzian and Protoaurignacian contexts. In particular, differences seem to emerge in the use of fire (especially in terms of temperatures and bone processing) between Late Mousterian layers and the subsequent phases. More substantial data on the distribution of ungulate limb elements suggest a marked change in prey exploitation between the Late Mousterian and the Early Upper

| TECHNOCOMPLEX | LM | | LM | | LM | | UL | | UL | | UL | | PA | | PA | |
|---|---|---|---|---|---|---|---|---|---|---|---|---|---|---|---|---|
| SITE - US/Lev. | OSC | | CTC | | CAV(all lev.) | | CAV EIII5 | | CTC | | CALA 14 | | CTC ars-gic | | CALA 13-10 | |
| TAXA | *Bos primig.* | | *Dama dama* | | *Bos primig.* | | *Bos primig.* | | *C. elaphus* | | *Dama dama* | | *C. elaphus* | | *C.. elaphus* | |
| | Nisp | % | Nisp | % | Nisp | % | Nisp | % | Nisp | % | Nisp | % | Nisp | % | Nisp | % |
| Antler/Horn | | | | | | | | | 3 | 3.2 | | | | | 1 | 0.2 |
| Cranium | 13 | 2.6 | 4 | 2.8 | 3 | 0.8 | 2 | 2.3 | 1 | 1 | 5 | 3.4 | | | 17 | 3.5 |
| Emimandible | 43 | 8.7 | 9 | 6.3 | 12 | 3 | | | 4 | 4.4 | 13 | 8.7 | 1 | 4.3 | 42 | 8.8 |
| Decidous teeth | 8 | 1.6 | 9 | 6.3 | 49 | 12.4 | 1 | 1.2 | 2 | 2.1 | | | 2 | 8.7 | | |
| Permanent teeth | 202 | 40.6 | 60 | 41.7 | 201 | 51 | 10 | 11.6 | 7 | 7.5 | | | 8 | 34.8 | | |
| Perm.+Deciduos teeth | 11 | 2.2 | | | | | | | | | 46 | 30.9 | | | 133 | 27.8 |
| Tooth indet. | 47 | 9.5 | 10 | 6.9 | 21 | 5.3 | 5 | 5.8 | | | 2 | 1.3 | 2 | 8.7 | 7 | 1.5 |
| Hioyd | | | | | 7 | 1.8 | 1 | 1.2 | | | | | | | | |
| Atlas-axis | | | | | | | | | | | | | | | | |
| Vertebra | | | | | | | | | 3 | 3.3 | 1 | 0.7 | | | 2 | 0.4 |
| Rib | | | | | | | | | 1 | 1 | | | | | | |
| Clavicle | | | | | | | | | | | | | | | | |
| Scapula | | | | | | | | | | | | | | | 1 | 0.2 |
| Humerus | 10 | 2 | 3 | 2.1 | | | 4 | 4.7 | 5 | 5.5 | 2 | 1.3 | | | 9 | 1.9 |
| Radius/Ulna | 5 | 1 | | | | | 2 | 2.3 | 1 | 1 | | | | | | |
| Radius | 10 | 2 | 5 | 3.5 | 9 | 2.3 | 1 | 1.2 | 4 | 4.4 | 6 | 4 | | | 12 | 2.5 |
| Ulna | 4 | 0.8 | | | 4 | 1 | 1 | 1.2 | | | 1 | 0.7 | | | 6 | 1.3 |
| Carpals | | | 2 | 1.4 | | | 4 | 4.7 | 8 | 8.7 | 3 | 2 | | | 15 | 3.1 |
| Metacarpal | 12 | 2.4 | 3 | 2.1 | 8 | 2 | 2 | 2.3 | 16 | 17.2 | | | 3 | 13 | 38 | 7.9 |
| Metacarpal rud. | | | | | | | | | | | | | | | | |
| Coxal | | | | | | | | | | | | | | | 1 | 0.2 |
| Femur | 5 | 1 | 4 | 2.8 | | | | | 4 | 4.4 | | | | | 2 | 0.4 |
| Patella | | | | | | | | | 1 | 1 | 1 | 0.7 | | | 1 | 0.2 |
| Tibia | 51 | 10.3 | | | 3 | 0.8 | 3 | 3.5 | 9 | 9.7 | 4 | 2.7 | | | 5 | 1 |
| Fibula | | | | | | | | | | | | | | | | |
| Malleolar bone | | | | | | | 1 | 1.2 | | | | | | | 3 | 0.6 |
| Calcaneum | | | | | | | | | | | | | | | | |
| Astragalus | | | | | | | | | | | | | | | | |
| Tarsals | 7 | 1.4 | 2 | 1.4 | 6 | 1.5 | 6 | 7 | | | 14 | 9.4 | | | 14 | 2.9 |
| Metatarsal | 38 | 7.6 | 12 | 8.3 | 21 | 5.3 | 5 | 5.8 | 14 | 15 | 22 | 14.7 | 3 | 13.2 | 78 | 16.3 |
| Metapodial | 4 | 0.8 | 3 | 2.1 | 16 | 4.1 | 2 | 2.3 | | | 8 | 5.4 | 2 | 8.7 | 24 | 5 |
| First phal. | 13 | 2.6 | 6 | 4.2 | 15 | 3.8 | 16 | 18.6 | 7 | 7.5 | 12 | 8.1 | 1 | 4.3 | 29 | 6.1 |
| Second phal. | 6 | 1.2 | 9 | 6.3 | 4 | 1 | 8 | 9.3 | 2 | 2.1 | 6 | 4 | 1 | 4.3 | 21 | 4.4 |
| Third phal. | 1 | 0.2 | 2 | 1.4 | 1 | 0.3 | 1 | 1.2 | | | | | | | 9 | 1.9 |
| First phal. rud. | | | | | | | | | | | | | | | | |
| Sec. phal. rud. | | | | | | | | | | | | | | | | |
| Th. phal. rud. | | | | | | | | | | | | | | | | |
| Sesamoid | 7 | 1.5 | 1 | 0.4 | 14 | 3.6 | 11 | 12.6 | 1 | 1 | 3 | 2 | | | 9 | 1.9 |
| TOTAL | 497 | | 144 | | 394 | | 86 | | 93 | | 149 | | 23 | | 479 | |
| Tot Cranium + tooth | *324* | *65.2* | *92* | *63.8* | *293* | *74.4* | *19* | *22.1* | *17* | *18.3* | *66* | *44.3* | *13* | *56.5* | *200* | *41.7* |
| Tot trunk | | | | | | | | | *4* | *4.3* | *1* | *0.7* | | | *4* | *0.9* |
| Tot long limb bones | *146* | *29.4* | *30* | *20.8* | *61* | *15.5* | *21* | *24.4* | *53* | *57* | *44* | *29.5* | *8* | *34.8* | *178* | *37.2* |
| Carpal + tarsal | | | *5* | *3.6* | *6* | *1.5* | *10* | *11.6* | *8* | *8.6* | *17* | *11.4* | | | *29* | *6* |
| Phal. + sesamoides | *27* | *5.4* | *17* | *11.8* | *34* | *8.6* | *36* | *41.9* | *11* | *11.8* | *21* | *14.1* | *2* | *8.7* | *68* | *14.2* |

**Tab. 7** Number of remains and relative % of specific anatomical elements referable to the most represented mammal species documented in the MP/UP transition layers and levels of Southern Italian sites (see Fig. 1 – Area 2+3). The subtotals of the different anatomical compartments are reported at the bottom of the table. CAV = Grotta del Cavallo; OSC = Riparo l'Oscurusciuto; CTC = Grotta di Castelcivita; CALA = Grotta della Cala.

Palaeolithic in southern Italy, while northern sites show that a higher variety of processing techniques was already present in the Late Mousterian. As concerns differences in hunting strategies, traces of an increasing preference for small- and medium-sized mammals (carnivores, rodents, lagomorphs) can be already documented for the transition to Protoaurignacian, although presently available evidence is exclusively qualitative. Future research will ascertain if this difference can be ascribed to a forced expansion of niche breadth due to economic and technological competition between Neandertals and modern humans (Hockett and Haws, 2005).

The above mentioned hypotheses cannot yet be tested because of small sample size in all the analysed classes, and the emerging trends may or may not be confirmed by adding evidence on the same sites as well as on other, currently underrepresented areas of the Italian Peninsula to the analyses presented here. Over the next three years, the project ERC n. 724046 - SUCCESS will build on the results presented here by acquiring novel zooarchaeological and chronological evidence on all the mentioned contexts (Fig. 1), by directly comparing faunal time series to palaeoenvironmental and palaeoclimatic data, and by relying on innovative methods (Pothier Bouchard et al., 2019; Pothier Bouchard et al., 2019) such as ZooMS (ZooArchaeology by Mass Spectrometry). This evidence will contribute to helping resolve or at least clarify longstanding debates surrounding strategic and technological shifts which occurred during the Middle-Upper Paleolithic transition and will help situate the questions concerning contacts between Neandertals and modern humans in Italy (and the eventual replacement of the former by the latter) in the broader framework of complex adaptive strategies and long-term human-environment interactions.

## Data availability

Datasets, scripts and related commands used to generate all of the results described in the paper are available at (http://doi.org/10.6092/unibo/amsacta/6209).

## Acknowledgments

This project has received funding from the European Research Council (ERC) under the European Union's Horizon 2020 research and innovation programme (grant agreement No 724046 - SUCCESS); website: http://www.ercsuccess.eu/. Research at Fumane is coordinated by the Ferrara University (M.P.) in the framework of a project supported by MIBAC, public institutions (Lessinia Mountain Community, Fumane Municipality and others). Research at Riparo del Broion (M.R.) and Grotta di San Bernardino (M.P.) is designed by Ferrara and Bologna University and was supported by MIBAC, the Province of Vicenza, the Veneto Region, the Italian Ministry of Research and Education, and institutions (Leakey Foundation, Spring 2015 Grant; Fondazione CariVerona). Research at Rio Secco (M.P. and M.R.) is designed by Ferrara University and was supported by MIBAC the Administration of the Clauzetto Municipality and the Friuli Venezia Giulia Region and a group of public institutions (Ecomuseo delle Dolomiti Friulane, "Lis Aganis", BIM Tagliamento Consortium, Pordenone Province), Foundations (Fondazione CRUP) and private companies (Friulovest Banca). Researches at Grotta di Castelcivita, Grotta della Cala and Riparo L'Oscurusciuto are coordinated by the University of Siena (A.R., A.M. and F.B. respectively). We thank the Soprintendenza Archeologia, Belle Arti e Paesaggio per le Province di Brindisi, Lecce e Taranto, and the Soprintendenza Archeologia, Belle Arti e Paesaggio per le Province di Salerno e Avellino for kindly supporting our research and fieldwork in Apulia and Campania over the years. We acknowledge the Municipalities of Camerota, Castelcivita, Ginosa and the Parco Nazionale del Cilento and Vallo di Diano for logistic support. Research at Riparo Bombrini is coordinated by the University of Genoa (F.N.) and Université de Montréal (J.R.S) and supported by FRQSC grant 2016-NP-193048 (J.R.S) and SSHRC Insight Grant 435-2017-1520 (J.R.S. & F.N.), under the auspices of the Soprintendenza Archeologia, Belle Arti e Paesaggio per la città metropolitana di Genova e le province di Imperia, La Spezia e Savona, with logistical support from the Istituto Internazionale di Studi Liguri (Bordighera), the Museo preistorico nazionale dei Balzi Rossi and the Polo museale della Liguria.

## Appendices.

| | RS 5+8 - LM | | RF A9 - LM | | RF A6 - LM | | RF A5/A5+A6 - LM | | RS 5top+7 - LM | | SB II+III - LM | | RF A4 - LM | | RB 1f-1g - UL | | RF A3 - UL | | RF A2-A2R - PA | |
|---|---|---|---|---|---|---|---|---|---|---|---|---|---|---|---|---|---|---|---|---|
| *Taxa* | Nisp | % | Nisp | % | Nisp | % | Nisp | % | Nisp | % | Nisp | % | Nisp | % | Nisp | % | Nisp | % | Nisp | % |
| *Stephanorhinus* sp. | | | | | | | | | | | 2 | 0.3 | | | | | | | | |
| *Coelodonta antiquitatis* | | | | | | | | | | | | | | | | | 1 | 0.2 | | |
| *Sus scrofa* | 2 | 4.8 | 2 | 0.2 | 2 | 0.1 | | | 1 | 1.7 | 36 | 5.2 | | | 21 | 35.6 | | | | |
| *Megaloceros giganteus* | 5 | 11.9 | 79 | 6.5 | 28 | 1.8 | 10 | 2.1 | 14 | 24.1 | 12 | 1.7 | 12 | 2.5 | 2 | 3.4 | 8 | 1.8 | 34 | 4.3 |
| *Cervus elaphus* | 7 | 16.7 | 495 | 40.8 | 1095 | 69.7 | 297 | 62 | 6 | 10.3 | 136 | 19.6 | 242 | 50 | 5 | 8.5 | 169 | 37.4 | 170 | 21.4 |
| *Capreolus capreolus* | 1 | 2.4 | 281 | 23.1 | 182 | 11.6 | 48 | 10 | | | 251 | 36.2 | 54 | 11.2 | 3 | 5.1 | 50 | 11.1 | 37 | 4.7 |
| *Alces alces* | 2 | 4.8 | 17 | 1.4 | 4 | 0.3 | 1 | 0.2 | 5 | 8.6 | 24 | 3.5 | | | 3 | 5.1 | | | | |
| Cervidae | 6 | 14.3 | 166 | 13.7 | 128 | 8.2 | 39 | 8.1 | 19 | 32.8 | 135 | 19.5 | 29 | 6 | 13 | 22 | 33 | 7.3 | | |
| *Bos primigenius* | | | 6 | 0.5 | | | | | | | 1 | 0.1 | | | 1 | 1.7 | | | 2 | 0.3 |
| *Bison priscus* | 1 | 2.4 | 6 | 0.5 | 2 | 0.1 | | | 2 | 3.4 | | | 5 | 1 | 1 | 1.7 | 5 | 1.1 | 6 | 0.8 |
| *Bos/Bison* | 10 | 23.8 | 29 | 2.4 | 13 | 0.8 | 10 | 2.1 | 8 | 13.8 | 33 | 4.8 | 16 | 3.3 | 1 | 1.7 | 24 | 5.3 | 28 | 3.5 |
| *Capra ibex* | 5 | 11.9 | 46 | 3.8 | 54 | 3.4 | 30 | 6.3 | 2 | 3.4 | 3 | 0.4 | 82 | 16.9 | 1 | 1.7 | 116 | 25.7 | 447 | 56.2 |
| *Rupicapra rupicapra* | 3 | 7.1 | 68 | 5.6 | 55 | 3.5 | 32 | 6.7 | | | 53 | 7.6 | 31 | 6.4 | 4 | 6.8 | 34 | 7.5 | 71 | 8.9 |
| Caprinae | | | 19 | 1.6 | 7 | 0.4 | 12 | 2.5 | 1 | 1.7 | 8 | 1.2 | 13 | 2.7 | 4 | 6.8 | 12 | 2.7 | | |
| Total Ungulata | 42 | | 1214 | | 1570 | | 479 | | 58 | | 694 | | 484 | | 59 | | 452 | | 795 | |

**Table A.1** Total Nisp and relative frequency of Ungulata documented in levels and layers of Northern Italy (Area 1) presented in chronological-cultural order. RF = Grotta di Fumane; RB = Riparo del Broion; SB = Grotta di San Bernardino; RS = Grotta del Rio Secco.

| | RS 5+8 - LM | | RF A9 - LM | | RF A6 - LM | | RF A5/A5+A6 - LM | | RS 5top+7 - LM | | SB II+III - LM | | RF A4 - LM | | RB 1e+1f+1g - UL | | RF A3 - UL | | RF A2-A2R - PA | |
|---|---|---|---|---|---|---|---|---|---|---|---|---|---|---|---|---|---|---|---|---|
| *Taxa* | Nisp | % | Nisp | % | Nisp | % | Nisp | % | Nisp | % | Nisp | % | Nisp | % | Nisp | % | Nisp | % | Nisp | % |
| *Canis lupus* | 3 | 1.3 | 4 | 11.1 | 7 | 11.9 | 4 | 7.7 | | | 3 | 1.8 | 11 | 11.5 | | | 21 | 25.6 | 38 | 37.3 |
| *Vulpes vulpes* | 3 | 1.3 | 6 | 16.7 | 20 | 33.9 | 26 | 50 | 1 | 0.5 | 9 | 5.5 | 61 | 63.5 | 4 | 9.5 | 36 | 43.9 | 28 | 27.5 |
| *Vulpes/alopex* | | | | | | | 3 | 5.8 | | | | | 5 | 5.2 | | | 6 | 7.3 | 7 | 6.9 |
| *Alopex lagopus* | | | | | | | | | | | | | | | | | | | 2 | 2 |
| *Ursus spelaeus* | 157 | 66.2 | 8 | 22.2 | | | 2 | 3.8 | 148 | 66.7 | 100 | 61.3 | 2 | 2.1 | 21 | 50 | | | 1 | 1 |
| *Ursus arctos* | 1 | 0.4 | 4 | 11.1 | 10 | 16.9 | 4 | 7.7 | 6 | 2.7 | 3 | 1.8 | 11 | 11.5 | | | 7 | 8.5 | 1 | 1 |
| *Ursus* sp. | 64 | 27 | 5 | 13.9 | 21 | 35.6 | 13 | 25 | 66 | 29.7 | 38 | 23.3 | 2 | 2.1 | 13 | 31 | 2 | 2.4 | 1 | 1 |
| *Mustela erminea* | 4 | 1.7 | | | | | | | | | | | | | | | 1 | 1.2 | 2 | 2 |
| *Mustela nivalis* | | | 4 | 11.1 | 1 | 1.7 | | | | | | | 1 | 1 | | | 2 | 2.4 | 1 | 1 |
| *Mustela putorius* | | | | | | | | | | | 2 | 1.2 | | | | | | | 1 | 1 |
| *Martes martes* | 3 | 1.3 | | | | | | | 1 | 0.5 | | | | | 1 | 2.4 | | | | |
| Mustelidae | | | 1 | 2. | | | | | | | | | | | | | | | | |
| *Gulo gulo* | | | | | | | | | | | | | 1 | 1 | | | 3 | 3.7 | 1 | 1 |
| *Meles meles* | 2 | 0.8 | | | | | | | | | | | | | | | | | | |
| *Crocuta crocuta spelaea* | | | 3 | 8.3 | | | | | | | | | | | | | 3 | 3.7 | 15 | 14.7 |
| *Felis silvestris* | | | | | | | | | | | 2 | 1.2 | | | 2 | 4.8 | | | | |
| *Lynx lynx* | | | | | | | | | | | 4 | 2.5 | | | | | 1 | 1.2 | 3 | 2.9 |
| *Panthera pardus* | | | | | | | | | | | 1 | 0.6 | 2 | 2.1 | | | | | | |
| *Panthera leo spelaea* | | | 1 | 2.8 | | | | | | | | | | | | | | | 1 | 1 |
| Felidae | | | | | | | | | | | 1 | 0.6 | | | 1 | 2.4 | | | | |
| Total Carnivora | 237 | | 36 | | 59 | | 52 | | 222 | | 163 | | 96 | | 42 | | 82 | | 102 | |

**Table A.2** Total Nisp and relative frequency of Carnivora documented in levels and layers of Northern Italy (Area 1 in Fig.1) presented in chronological-cultural order. RF = Grotta di Fumane; RB=Riparo del Broion; SB=Grotta di San Bernardino; RS=Grotta del Rio Secco.

| | RS 5+8-LM | RF A9-LM | RF A6-LM | RF A5/A5+A6-LM | RS 5top+7-LM | SB II+III-LM | RF A4-LM | RB 1e+1f+1g-UL | RF A3-UL | RF A2-A2R-PA |
|---|---|---|---|---|---|---|---|---|---|---|
| *Taxa* | Nisp | Nisp | Nisp | Nisp | Nisp | Nisp | Nisp | Nisp | Nisp | Nisp |
| *Marmota marmota* | | 8 | 1 | | | 18 | | 3 | 2 | 2 |
| *Lepus* cfr. *timidus* | | | | | | | | 2 | | 4 |
| *Lepus* sp. | 1 | | | 1 | | 3 | | | 3 | 1 |
| *Castor fiber* | | | | | | 27 | | 1 | | 1 |
| Total Lagomorpha and Rodentia | 1 | 8 | 1 | 1 | 0 | 48 | 0 | 6 | 5 | 8 |

**Table A.3** Total Nisp and relative frequency of Rodentia and Lagomorpha documented in levels and layers of Northern Italy (Area 1 in Fig.1) presented in chronological-cultural order. RF = Grotta di Fumane; RB = Riparo del Broion; SB = Grotta di San Bernardino; RS = Grotta del Rio Secco.

| | CTC spits 30-33 LM | | CTC spits 25-29 LM | | CTC spits 21-24 LM | | CTC spit 20 LM | | CTC 18lower-19 LM | | CTC spit 18upper UL | | CTC spit 17-3 UL | | CTC spits 12-10 lower UL | | CALA 14 UL | | CTC spits 10 upper-8 PA | | CTC spit 7-top of seq.PA | | CALA 13 PA | | CALA 12 PA | | CALA 11-10 PA | |
|---|---|---|---|---|---|---|---|---|---|---|---|---|---|---|---|---|---|---|---|---|---|---|---|---|---|---|---|---|
| *Taxa* | Nisp | % | Nisp | % | Nisp | % | Nisp | % | Nisp | % | Nisp | % | Nisp | % | Nisp | % | Nisp | % | Nisp | % | Nisp | % | Nisp | % | Nisp | % | Nisp | % |
| *Stephanorhinus* sp. | | | | | 1 | 2.4 | | | | | | | | | | | 1 | 0.3 | | | | | | | | | | |
| *Equus ferus* | | | 1 | 0.5 | 1 | 2.4 | 1 | 2 | | | 1 | 5.9 | 7 | 6 | 65 | 59.1 | 18 | 5.2 | 10 | 30.3 | 1 | 1.7 | 5 | 2.2 | 10 | 2.3 | 4 | 1.8 |
| *Sus scrofa* | 2 | 1.5 | 4 | 2.1 | 1 | 2.4 | | | | | | | 4 | 3.4 | 4 | 3.6 | 46 | 13.3 | 7 | 21.2 | 4 | 6.7 | 10 | 4.3 | 52 | 12.1 | 28 | 12.3 |
| *Cervs elaphus* | 28 | 21.4 | 49 | 26.2 | 16 | 39 | 8 | 16 | 6 | 13.6 | 1 | 5.9 | 33 | 28.2 | 15 | 13.6 | 58 | 16.7 | 6 | 18.2 | 26 | 43.3 | 132 | 57.4 | 213 | 49.8 | 134 | 58.8 |
| *Capreolus capreolus* | 4 | 3.1 | 8 | 4.3 | 4 | 9.8 | 8 | 16 | 5 | 11.4 | 5 | 29.4 | 24 | 20.5 | 5 | 4.5 | 34 | 9.8 | 2 | 6.1 | 8 | 13.3 | 8 | 3.5 | 44 | 10.3 | 17 | 7.5 |
| *Dama dama* | 51 | 38.9 | 63 | 33.7 | 11 | 26.8 | 5 | 10 | 7 | 15.9 | 1 | 5.9 | 16 | 13.7 | 3 | 2.7 | 152 | 43.8 | 1 | 3 | 1 | 1.7 | 42 | 18.3 | 62 | 14.5 | 18 | 7.9 |
| Cervidae indet. | 6 | 4.6 | 3 | 1.6 | | | 8 | 16 | | | 1 | 5.9 | 3 | 2.6 | 1 | 0.9 | 13 | 3.7 | | | 3 | 5 | 10 | 4.3 | 22 | 5.1 | 9 | 3.9 |
| *Bos primigenus* | | | | | | | | | | | | | | | | | 16 | 4.6 | | | | | 7 | 3 | 8 | 1.9 | 2 | 0.9 |
| *Bison priscus* | 2 | 1.5 | 7 | 3.7 | | | 1 | 2 | | | 1 | 5.9 | 3 | 2.6 | 5 | 4.5 | | | 6 | 18.2 | 1 | 1.7 | | | | | | |
| *Bos/Bison* | | | | | | | 2 | 4 | 2 | 4.5 | 1 | 5.9 | 3 | 2.6 | 4 | 3.6 | | | | | | | | | | | | |
| *Capra ibex* | 13 | 9.9 | 34 | 18.2 | 2 | 4.9 | 2 | 4 | 2 | 4.5 | 1 | 5.9 | 5 | 4.3 | 6 | 5.5 | 3 | 0.9 | 1 | 3 | 3 | 5 | 12 | 5.2 | 14 | 3.3 | 15 | 6.6 |
| *Rupicapra* sp | 25 | 19.1 | 18 | 9.6 | 5 | 12.2 | 15 | 30 | 22 | 50 | 5 | 29.4 | 19 | 16.2 | 1 | 0.9 | 4 | 1.2 | | | 13 | 21.7 | 4 | 1.7 | 2 | 0.5 | 1 | 0.4 |
| Caprinae | | | | | | | | | | | | | | | | | 2 | 0.6 | | | | | | | 1 | 0.2 | | |
| Total Ungulata | 131 | | 187 | | 41 | | 50 | | 44 | | 17 | | 117 | | 110 | | 347 | | 33 | | 60 | | 230 | | 428 | | 228 | |

**Table A.4** Total Nisp and relative frequency of Ungulata documented in levels and layers of Southwestern-Tyrrhenian Italy (Area 2 in Fig.1) presented in chronological-cultural order. CTC = Grotta di Castelcivita; CALA = Grotta della Cala.

| | CTC gar LM | CTC lower rsi LM | CTC spit 18 upper UL | CTC spits 17-13 UL | CTC spits 12–10 lower UL | CALA 14 UL | CTC PA | CALA 13 PA | CALA 12 PA | CALA 11 PA | CALA 10 PA |
|---|---|---|---|---|---|---|---|---|---|---|---|
| *Taxa* | Nisp | Nisp | Nisp | Nisp | Nisp | Nisp | Nisp | Nisp | Nisp | Nisp | Nisp |
| *Canis lupus* | | | | | 1 | | 2 | 2 | 1 | | |
| *Vulpes vulpes* | | | | 2 | 2 | 9 | | 2 | 1 | | |
| *Ursus spelaeus* | | 1 | | | 1 | | | | | | |
| *Ursus arctos* | | | | 4 | | 7 | | | 2 | 1 | |
| *Mustela nivalis* | 1 | | | 1 | 2 | | | | | | |
| *Martes* sp. | | | | | | 14 | | | 1 | | |
| Mustelidae | | | | 2 | | | | | | | |
| *Meles meles* | | | | 2 | | | | | | | |
| *Crocuta crocuta spelaea* | 11 | | | 1 | 6 | | 1 | | | | |
| *Felis silvestris* | | | | 2 | 3 | 5 | | 2 | | | |
| *Panthera pardus* | 3 | 1 | | 3 | 2 | 17 | | 4 | 6 | 3 | |
| *Panthera leo spel.* | | | | | | | | | | | |
| Carnivora indet. | 6 | 5 | 1 | 7 | 1 | 3 | 2 | 1 | 1 | 1 | |
| Total Carnivora | 21 | 7 | 1 | 24 | 18 | 55 | 5 | 11 | 12 | 5 | 0 |

**Table A.5** Total Nisp and relative frequency of Carnivora documented in levels and layers of Southwestern-Tyrrhenian Italy (Area 2 in Fig.1) presented in chronological-cultural order. CTC = Grotta di Castelcivita; CALA = Grotta della Cala.

| Taxa | CAV FIIIeb-e LM Nisp | % | CAV FIIIb-c-d LM Nisp | % | CAV FI-II-IIIa LM Nisp | % | OSC 4-13 LM Nisp | % | OSC 3 LM Nisp | % | OSC 2-29-30-31 LM Nisp | % | OSC 1 LM Nisp | % | CAV EIII 5 UL Nisp | % |
|---|---|---|---|---|---|---|---|---|---|---|---|---|---|---|---|---|
| *Stephanorhinus sp.* | | | | | | | 1 | 0.2 | 7 | 12.3 | | | | | | |
| *Equus ferus* | 40 | 11.5 | 40 | 14.9 | 50 | 19.8 | 17 | 3 | 16 | 28.1 | 48 | 25.9 | 2 | 5 | 53 | 27.3 |
| *Equus hydruntinus* | | | | | | | | | | | | | | | 1 | 0.5 |
| *Equus sp.* | | | | | | | | | | | | | | | 1 | 0.5 |
| *Sus scrofa* | 1 | 0.3 | 4 | 1.5 | 2 | 0.8 | 1 | 0.2 | | | | | | | 1 | 0.5 |
| *Cervus elaphus* | 72 | 20.6 | 54 | 20.1 | 69 | 27.3 | 51 | 8.9 | 15 | 26.3 | 27 | 14.6 | 11 | 27.5 | 52 | 26.8 |
| *Capreolus capreolus* | | | 6 | 2.2 | 3 | 1.2 | 11 | 1.9 | 2 | 3.5 | 8 | 4.3 | 1 | 2.5 | | |
| *Dama dama* | 7 | 2 | 83 | 31 | 20 | 7.9 | 38 | 6.6 | 1 | 1.8 | 12 | 6.5 | 6 | 15 | | |
| *Cervidae indet.* | 2 | 0.6 | 15 | 5.6 | 8 | 3.2 | 6 | 1 | | | 3 | 1.6 | | | | |
| *Bos primigenus* | 227 | 65 | 66 | 24.6 | 101 | 39.9 | 445 | 77.5 | 15 | 26.3 | 82 | 44.3 | 20 | 50 | 86 | 44.3 |
| *Capra ibex* | | | | | | | 1 | 0.2 | 1 | 1.8 | 5 | 2.7 | | | | |
| *Rupicapra sp.* | | | | | | | 3 | 0.5 | | | | | | | | |
| Total Nisp | 349 | | 268 | | 253 | | 574 | | 57 | | 185 | | 40 | | 194 | |

**Table A.6** Total Nisp and relative frequency of Ungulata documented in levels and layers of Southeastern (Ionian-Adriatic) Italy (Area 3 in Fig.1) presented in chronological-cultural order. CAV = Grotta del Cavallo; OSC = Riparo l'Oscurusciuto.

| | CAV FIII LM | CAV F II LM | OSC US 4-13 LM | OSC US 3 LM | OSC US 2-29-31 LM | OSC US 1 LM | CAV EIII5 UL |
|---|---|---|---|---|---|---|---|
| *Taxa* | Nisp | Nisp | Nisp | Nisp | Nisp | Nisp | Nisp |
| *Canis lupus* | | | 1 | | | | 2 |
| *Vulpes vulpes* | 42 | 13 | | | | | 4 |
| *Ursus spelaeus* | | | | | | | |
| *Ursus arctos* | | | | | | | |
| *Mustela nivalis* | | | | | | | |
| *Martes* sp. | | | | | | | |
| *Mustelidae* | | | | | | | |
| *Meles meles* | | | | | | | |
| *Crocuta crocuta spelaea* | | | | | | | 1 |
| *Felis silvestris* | 2 | | | | | | |
| *Panthera pardus* | | | | | | | |
| *Panthera leo spelaea* | | | | | 1 | | |
| Carnivora indet. | 2 | | | | | | |
| Total Carnivora | 46 | 13 | 1 | 0 | 1 | 0 | 7 |

**Table A.7** Total Nisp and relative frequency of Carnivora documented in levels and layers of Southeastern (Ionian-Adriatic) Italy (Area 3 in Fig.1) presented in chronological-cultural order. CAV = Grotta del Cavallo; OSC = Riparo l'Oscurusciuto.

| | FUMANE | | | | | | | | | | | |
|---|---|---|---|---|---|---|---|---|---|---|---|---|
| | A9-LM | | A6-LM | | A5/A5+A6-LM | | A4-LM | | A3-UL | | A1-2 PA | |
| | NISP | % | NISP | % | NISP | % | NISP | % | NISP | % | NISP | % |
| *Coturnix coturnix* | 3 | 0.6 | 1 | 0.9 | 1 | 0.6 | | | 2 | 1.6 | | |
| *Alectoris graeca* | 2 | 0.4 | | | | | 1 | 0.7 | | | | |
| *Perdix perdix* | 2 | 0.4 | | | | | 3 | 2.1 | 1 | 0.8 | 1 | 0.4 |
| *Lagopus* cf. *lagopus* | | | 1 | 0.9 | | | | | | | | |
| *Lagopus muta* | | | 1 | 0.9 | | | | | 1 | 0.8 | 3 | 1.2 |
| *T. urogallus/L. tetrix tetr.* | 2 | 0.4 | | | | | | | | | | |
| *Lyrurus tetrix* | 24 | 5.1 | 8 | 6.9 | 22 | 12.5 | 28 | 19.4 | 24 | 18.9 | 45 | 18.3 |
| cf. *Lyrurus tetrix* | | | 3 | 2.6 | 6 | 3.4 | 2 | 1.4 | 2 | 1.6 | | |
| Galliformes unid. | 1 | 0.2 | | | 3 | 1.7 | | | | | | |
| *Aythya nyroca* | | | | | | | | | | | | |
| *Spatula querquedula* | | | | | | | | | | | | |
| *Mareca strepera* | | | | | | | | | | | | |
| *Mareca penelope* | | | | | | | | | | | | |
| *Anas platyrhynchos* | | | | | | | 1 | 0.7 | | | | |
| *Anas crecca* | | | | | | | | | | | | |
| *Anas* cf. *crecca* | 1 | 0.2 | | | | | | | | | | |
| *Columba livia/oenas* | 2 | 0.4 | | | | | | | | | | |
| *Columba oenas* | | | | | | | | | | | 1 | 0.4 |
| *Columba palumbus* | | | 1 | 0.9 | | | | | | | | |
| *Caprimulgus europaeus* | | | | | | | | | | | | |
| *Rallus aquaticus* | 2 | 0.4 | | | 1 | 0.6 | 1 | 0.7 | | | 1 | 0.4 |
| *Crex crex* | 70 | 14.7 | 16 | 13.8 | 24 | 13.6 | 24 | 16.7 | 25 | 19.7 | 53 | 21.5 |
| cf. *Crex crex* | | | 1 | 0.9 | 6 | 3.4 | | | | | | |
| cf. *Gallinula chloropus* | | | 1 | 0.9 | | | | | | | | |

| | CASTELCIVITA | | | | | | | | | | | | | |
|---|---|---|---|---|---|---|---|---|---|---|---|---|---|---|
| | rsi lower-LM | | Pie-UL | | Rpi-UL | | rsa'-UL | | rsa'-PA | | Gic-PA | | ars-PA | |
| | NISP | % | NISP | % | NISP | % | NISP | % | NISP | % | NISP | % | NISP | % |
| *Coturnix coturnix* | 1 | 2.3 | 5 | 3.3 | 2 | 2.4 | | | | | 2 | 25 | | |
| *Alectoris graeca* | 5 | 11.6 | 24 | 15.8 | 20 | 23.5 | 6 | 23.1 | 19 | 24.4 | 1 | 12.5 | 10 | 43.5 |
| *Perdix perdix* | 5 | 11.6 | 45 | 29.6 | 30 | 35.3 | 11 | 42.3 | 11 | 14.1 | 1 | 12.5 | 5 | 21.7 |
| *Lagopus* cf. *lagopus* | | | | | | | | | | | | | | |
| *Lagopus muta* | | | | | | | | | | | | | | |
| *T. urogallus/L. tetrix tetr.* | | | | | | | | | | | | | | |
| *Lyrurus tetrix* | | | | | | | | | | | | | | |
| cf. *Lyrurus tetrix* | | | | | | | | | | | | | | |
| Galliformes unid. | | | | | | | | | | | | | | |
| *Aythya nyroca* | | | 2 | 1.3 | | | 2 | 7.7 | | | | | | |
| *Spatula querquedula* | 1 | 2.3 | 13 | 8.6 | | | | | | | | | | |
| *Mareca strepera* | 5 | 11.6 | | | | | | | 1 | 1.3 | | | | |
| *Mareca penelope* | 1 | 2.3 | | | | | | | | | | | | |
| *Anas platyrhynchos* | | | 1 | 0.7 | | | | | | | | | | |
| *Anas crecca* | | | | | | | | | 2 | 2.6 | | | 1 | 4.3 |
| *Anas* cf. *crecca* | | | | | | | | | | | | | | |
| *Columba livia/oenas* | | | | | | | | | | | | | | |
| *Columba oenas* | 2 | 4.7 | 11 | 7.2 | 9 | 10.6 | 1 | 3.8 | 6 | 7.7 | | | 2 | 8.7 |
| *Columba palumbus* | | | | | | | | | | | | | | |
| *Caprimulgus europaeus* | | | | | | | | | | | 1 | 12.5 | | |
| *Rallus aquaticus* | | | | | | | | | | | | | | |
| *Crex crex* | | | 4 | 2.6 | 2 | 2.4 | | | 2 | 2.6 | | | 1 | 4.3 |
| cf. *Crex crex* | | | | | | | | | | | | | | |
| cf. *Gallinula chloropus* | | | | | | | | | | | | | | |

| | | | | | | | | | | | | | | | | | | | | | | | | | | |
|---|---|---|---|---|---|---|---|---|---|---|---|---|---|---|---|---|---|---|---|---|---|---|---|---|---|---|
| Rallidae unid. | 13 | 2.7 | 2 | 1.7 | 1 | 0.6 | | | | | | | | | | | | | | | | | | | | |
| *Burhinus oedicnemus* | | | | | | | | | | | | | | | | | 1 | 1.2 | | | | | | | | |
| *Pluvialis squatarola* | | | | | | | | | | | | | | | | | | | | | 4 | 5.1 | | | | |
| *Vanellus vanellus* | | | | | 1 | 0.6 | | | | | 1 | 0.4 | | | | | | | | | | | | | | |
| *Numenius phaeopus* | | | | | | | | | | | | | | | | | | | | | 1 | 1.3 | | | | |
| *Limosa limosa* | | | | | | | | | | | | | 2 | 4.7 | | | | | | | | | | | | |
| *Arenaria interpres* | | | | | | | | | | | | | | | 1 | 0.7 | | | | | | | | | | |
| *Calidris pugnax* | | | | | | | | | | | | | | | | | 1 | 1.2 | | | | | | | | |
| *Scolopax rusticola* | | | | | 1 | 0.6 | 1 | 0.7 | | | 1 | 0.4 | | | | | | | | | | | | | | |
| *Gallinago media* | | | | | | | | | | | | | | | 1 | 0.7 | | | | | | | | | | |
| *Larus ridibundus* | | | | | | | | | | | | | | | 1 | 0.7 | | | | | | | 1 | 12.5 | 1 | 4.3 |
| *Athene noctua* | | | | | | | | | | | | | | | 1 | 0.7 | | | | | | | | | | |
| *Aegolius funereus* | | | 2 | 1.7 | | | | | | | | | | | | | | | | | | | | | | |
| *Otus scops* | | | | | | | | | | | | | | | | | 1 | 1.2 | | | | | | | | |
| *Asio otus* | 5 | 1.1 | 1 | 0.9 | 3 | 1.7 | 6 | 4.2 | 1 | 0.8 | 10 | 4.1 | 1 | 2.3 | 6 | 3.9 | 4 | 4.7 | | | | | | | | |
| *Asio flammeus* | | | | | | | 2 | 1.4 | 2 | 1.6 | | | | | | | | | | | | | | | | |
| *Asio* cf. *flammeus* | 3 | 0.6 | | | | | | | | | | | | | | | | | | | | | | | | |
| *Asio* sp. | 1 | 0.2 | 1 | 0.9 | 1 | 0.6 | 2 | 1.4 | 2 | 1.6 | | | | | | | | | | | | | | | | |
| *Strix aluco* | | | | | | | | | | | 2 | 0.8 | | | | | 1 | 1.2 | | | 4 | 5.1 | | | 2 | 8.7 |
| cf. *Strix aluco* | | | | | | | | | 1 | 0.8 | | | | | | | | | | | | | | | | |
| *Gypaetus barbatus* | | | 1 | 0.9 | | | | | | | | | | | | | | | | | | | | | | |
| cf. *Gypaetus barbatus* | 1 | 0.2 | | | | | | | | | | | | | | | | | | | | | | | | |
| *Aegypius monachus* | 1 | 0.2 | 1 | 0.9 | | | | | | | | | | | | | | | | | | | | | | |
| cf. *Aegypius monachus* | 1 | 0.2 | | | | | | | | | | | | | | | | | | | | | | | | |
| *Clanga clanga* | 1 | 0.2 | | | | | | | | | | | | | | | | | | | | | | | | |
| *Aquila chrysaetos* | | | | | | | 1 | 0.7 | | | | | | | | | | | | | | | | | | |
| *Buteo lagopus* | | | | | | | | | | | 1 | 0.4 | | | | | | | | | | | | | | |
| *Buteo buteo* | | | | | | | | | | | | | | | | | | | | | 1 | 1.3 | | | | |
| *Coracias garrulus* | | | | | | | | | | | | | | | 2 | 1.3 | 2 | 2.4 | | | | | | | | |
| *Dendrocopos leucotos* | | | | | | | | | | | 1 | 0.4 | | | | | | | | | | | | | | |
| *Falco naumanni* | | | | | | | | | | | | | | | | | | | 1 | 3.8 | | | | | | |
| *Falco tinnunculus* | | | 4 | 3.4 | 7 | 4 | | | 4 | 3.1 | 6 | 2.4 | | | | | | | | | | | | | | |
| *Falco* cf. *tinnunculus* | 3 | 0.6 | | | | | | | | | | | | | | | | | | | | | | | | |
| *Falco vespertinus* | 2 | 0.4 | 1 | 0.9 | 4 | 2.3 | | | | | 1 | 0.4 | | | | | | | | | | | | | | |
| *Falco vespert./columb.* | 3 | 0.6 | | | | | | | | | | | | | | | | | | | | | | | | |
| *Falco columbarius* | 1 | 0.2 | | | | | | | | | | | | | | | | | | | | | | | | |
| *Falco subbuteo* | | | 1 | 0.9 | 3 | 1.7 | 1 | 0.7 | 1 | 0.8 | 6 | 2.4 | | | 2 | 1.3 | 2 | 2.4 | 2 | 7.7 | | | | | | |
| *Falco* sp. | | | | | | | 1 | 0.7 | | | | | | | | | | | | | | | | | | |
| *Pyrrhocorax pyrrhocorax* | 1 | 0.2 | | | | | 1 | 0.7 | | | | | 5 | 11.6 | 3 | 2 | | | 1 | 3.8 | 4 | 5.1 | | | | |
| *Pyrrhocorax graculus* | 166 | 34.9 | 39 | 33.6 | 64 | 36.4 | 44 | 30.6 | 48 | 37.8 | 96 | 39 | 11 | 25.6 | 17 | 11.2 | 7 | 8.2 | | | 14 | 17.9 | | | | |
| cf. *Pyrrhocorax graculus* | | | 6 | 5.2 | 21 | 11.9 | 5 | 3.5 | 5 | 3.9 | | | 1 | 2.3 | | | | | | | | | | | | |
| *P.graculus/C.monedula* | 6 | 1.3 | | | | | | | | | | | | | | | | | | | | | | | | |
| *Pyrrhocorax* sp. | 10 | 2.1 | | | | | | | | | | | | | | | | | | | | | | | | |
| *Garrulus glandarius* | 5 | 1.1 | | | | | | | | | 1 | 0.4 | | | | | | | | | | | | | | |
| *Pica pica* | 2 | 0.4 | | | 3 | 1.7 | 2 | 1.4 | 1 | 0.8 | 4 | 1.6 | | | | | | | | | | | | | | |
| *Nucifraga caryocatactes* | | | | | | | | | 1 | 0.8 | 1 | 0.4 | | | 5 | 3.3 | | | 1 | 3.8 | 1 | 1.3 | | | | |
| *Corvus monedula* | 2 | 0.4 | | | | | 1 | 0.7 | | | 2 | 0.8 | | | | | | | | | | | | | | |
| *Corvus corax* | | | | | 1 | 0.6 | | | | | | | | | | | | | | | | | | | | |
| *Corvus* cf. *corax* | 1 | 0.2 | | | | | | | | | | | | | | | | | | | | | | | | |
| *Corvus corone* | | | | | | | 1 | 0.7 | | | | | | | | | | | | | | | | | | |
| *Corvus* cf. *corone* | 1 | 0.2 | | | | | | | | | | | | | | | | | | | | | | | | |
| Corvidae unid. | 11 | 2.3 | 4 | 3.4 | 3 | 1.7 | 3 | 2.1 | 1 | 0.8 | | | | | | | | | | | | | | | | |
| *Lullula arborea* | | | | | | | | | | | 1 | 0.4 | | | | | | | | | | | | | | |
| *Delichon urbicum* | | | | | | | | | | | | | 1 | 2.3 | | | | | | | 5 | 6.4 | | | | |
| *Ptyonoprogne rupestris* | 4 | 0.8 | 1 | 0.9 | | | 1 | 0.7 | | | 1 | 0.4 | | | 4 | 2.6 | 1 | 1.2 | | | 3 | 3.8 | | | | |
| *Turdus viscivorus* | 2 | 0.4 | | | | | | | | | 1 | 0.4 | 2 | 4.7 | 2 | 1.3 | 2 | 2.4 | 1 | 3.8 | | | 2 | 25 | 1 | 4.3 |
| *Turdus visciv/pilaris* | 1 | 0.2 | | | | | | | | | | | | | | | | | | | | | | | | |
| *Turdus iliacus* | | | | | | | | | | | | | | | 2 | 1.3 | | | | | | | | | | |
| *Turdus pilaris* | 1 | 0.2 | | | | | | | | | 2 | 0.8 | | | | | | | | | | | | | | |
| *Turdus* sp. | 1 | 0.2 | | | | | | | | | | | | | | | | | | | | | | | | |
| *Montifringilla nivalis* | | | 1 | 0.9 | | | | | 1 | 0.8 | 1 | 0.4 | | | | | | | | | | | | | | |
| *Pyrrhula pyrrhula* | 3 | 0.6 | | | | | | | 1 | 0.8 | | | | | | | | | | | | | | | | |
| *Linaria cannabina* | | | | | | | 1 | 0.7 | | | | | | | | | | | | | | | | | | |
| *Loxia pytyopsittacus* | | | | | | | | | | | 3 | 1.2 | | | | | | | | | | | | | | |
| Passeriformes unid. | 114 | 24 | 18 | 15.5 | | | 11 | 7.6 | 3 | 2.4 | | | | | | | | | | | | | | | | |
| Total Nisp | 475 | | 116 | | 176 | | 144 | | 127 | | 246 | | 43 | | 152 | | 85 | | 26 | | 78 | | 8 | | 23 | |

**Table A.8** Nisp and relative frequency of avifaunal taxa recovered in different levels and layers of Grotta di Fumane and Grotta di Castelcivita, in chronological-cultural order.

Click here to access/download
**Table**
Table A9.xls

**Table A.9** The most recently available dates for the context of interest used in the present work.

| Sites levels - US | Single teeth | | Carpal and tarsal bones | | Phalanges + sesamoides | | Total ungulates |
|---|---|---|---|---|---|---|---|
| | NR | % | NR | % | NR | % | Nisp |
| CALA PA | 299 | 35.4 | 188 | 22.3 | 115 | 13.6 | 844 |
| CTC PA | 2 | 5.6 | 4 | 11.1 | 5 | 13.9 | 38 |
| CALA UL | 137 | 41.5 | 31 | 9.4 | 41 | 12.4 | 331 |
| CAV EIII5 UL | 46 | 23.5 | 30 | 15.3 | 59 | 30.1 | 196 |
| CTC UL | 38 | 15.1 | 13 | 5.2 | 38 | 15.1 | 233 |
| OSC US 2 LM | 128 | 69.2 | 3 | 1.6 | 5 | 2.7 | 185 |
| CAV str. F LM | 552 | 65.3 | 12 | 1.4 | 67 | 7.9 | 845 |
| CTC LM | 6 | 9 | 3 | 4.5 | 9 | 13.4 | 67 |

**Table A.10** Number and % of single teeth and small limb bones of Ungulata uncovered in different levels of the Late Mousterian, Uluzzian and Protoaurignacian sites of Southern Italy. CALA = Grotta della Cala; CTC = Grotta di Castelcivita; CAV = Grotta del Cavallo; OSC = Riparo l'Oscurusciuto.

| | OSC US 4/1 LM | CAV FII LM | CTC LM | CAV EIII5 UL |
|---|---|---|---|---|
| Elements | % | % | % | % |
| | | | | |
| Antler/Horn | 0.4 | 0.3 | | 3.5 |
| Skull | 4.3 | 4.8 | 4.9 | 2.5 |
| Mandible | 0.4 | 0.4 | 2 | 0.7 |
| Teeth | 18.4 | 14.9 | 3.6 | 7.9 |
| Vertebrae | | 3 | 4.6 | 3.3 |
| Ribs | 3.5 | 11.2 | 15.7 | 13.4 |
| Scapula | | 2.3 | | 0.4 |
| Sternum | | | 0.7 | 0.7 |
| Pelvis | 0.4 | | 0.3 | 0.2 |
| Metapodials | | | 0.8 | |
| Diaphysis | 45.3 | 41.3 | 41.4 | 18.6 |
| Epiphysis | 5.6 | 8.2 | 8.2 | 12.9 |
| Spongy bones | 21.8 | 11.6 | 12 | 31.9 |
| Total remains | 5747 | 9574 | 1920 | 5185 |

**Table A.11** Percentages of determinated skeletal parts in the taxonomically indeterminate remains recovered from the different Late Mousterian and Uluzzian layers and levels sampled in Southern Italy. OSC = Riparo l'Oscurusciuto; CAV = Grotta del Cavallo; CTC = Grotta di Castelcivita.

| Sites | US/levels | GM | (D)igested | TOT.Car.M | TOT % | TOT. NR |
|---|---|---|---|---|---|---|
| RS-Rio Secco | 5+8 | 53 | 2 | 55 | 1.3 | 4301 |
| RF-Fumane | A9 | 100 | 1 | 101 | 0.09 | 111841 |
| RF-Fumane | A6 | 24 | 16 | 40 | 0.03 | 111044 |
| RF-Fumane | A5/A5+A6 | 20 | 9 | 29 | 0.04 | 67083 |
| RS-Rio Secco | 5top+7 | 31 | - | 31 | 5.9 | 524 |
| SB-S. Bernardino | II+III | 61 | 1 | 62 | 0.6 | 9217 |
| RF-Fumane | A4 | 51 | 17 | 68 | 0.3 | 19955 |
| RB-Broion | 1e+1f+1g | 3 | 1 | 4 | 0.01 | 37390 |
| RF-Fumane | A3 | 53 | 36 | 89 | 0.5 | 16989 |
| RF-Fumane | A2-A2R | 17 | 9 | 26 | 0.1 | 19829 |

**Table A.12** Detail and percentages of remains that present with evidence of digestion and/or gnawing by carnivores documented in the levels and layers of e Northern Italy (Adriatic Area 1 in Fig. 1). Contexts are presented in chronological-cultural order Fig. 1. GM: gnawing marks; TOT CM: total carnivore marks; D: digested.

| | Mann-Whitney W | P-value |
|---|---|---|
| 1-3cm LM – UL Northern Italy | 7 | 1 |
| >3cm LM – UL Northern Italy | 4 | 0.5 |
| Burn.+Calc LM – UL Northern Italy | 5 | 0.86 |

**Table A.13** Results of Mann-Whitney test for assessing significant differences in the distribution of fragment size classes and the proportion of burned and calcinated remains across Uluzzian and Late Mousterian layers of Northern Italy. The test was run on arcsine-transformed proportions.

| | X-squared | df | P-value | Effect size (Cohen's h) | Power |
|---|---|---|---|---|---|
| 1-3cm CAV UL – CAV LM | 128,7 | 1 | <0.001 | -0,19 | 1 |
| 1-3cm CAV UL – OSC LM | 1875,8 | 1 | <0.001 | -0,55 | 1 |
| >3cm CAV UL – CAV LM | 128,7 | 1 | <0.001 | 0,19 | 1 |
| >3cm CAV UL – OSC LM | 1875,8 | 1 | <0.001 | 0,58 | 1 |
| Burn+Calc CAV UL – CAV LM | 4264,7 | 1 | <0.001 | 1,1 | 1 |
| Burn+Calc CAV UL – OSC LM | 2,4161 | 1 | 0,12 | -0,04 | 0,72 |

**Table A.14** Results of test for differences in proportion of fragment size classes between Uluzzian and Late Mousterian layers of southeastern Italy (i.e. those with no direct evidence of carnivore gnawing) with relative effect size and statistical power.

| | X-squared | df | P-value | Effect size (Cohen's h) | Power |
|---|---|---|---|---|---|
| Carpal+tarsal CAV UL – CAV LM | 19,344 | 1 | <0.001 | 0,45 | 0,965 |
| Carpal+tarsal CAV UL – OSC LM | 52,104 | 1 | <0.001 | 0,69 | 0,999 |
| Phalang.+Sesamoides CAV UL – CAV LM | 59,942 | 1 | <0.001 | 0,81 | 0,999 |
| Phalang.+Sesamoides CAV UL – OSC LM | 97,192 | 1 | <0.001 | 0,93 | 1 |

**Table A.15** Results of test for differences in proportion of carpal/tarsal and phalanges/sesamoides of *Bos primigenius* between Uluzzian and Late Mousterian layers of southeastern Italy (i.e. those with no direct evidence of carnivore gnawing) with the relative effect size and statistical power.

| | X-squared | df | P-value | Effect size (Cohen's h) | Power |
|---|---|---|---|---|---|
| Carpal+tarsal CAV UL – CAV LM | 79,232 | 1 | <0.001 | 0,57 | 0,999 |
| Carpal+tarsal CAV UL – OSC LM | 20,831 | 1 | <0.001 | 0,55 | 0,999 |
| Phalanges+Sesamoides CAV UL – CAV LM | 73,523 | 1 | <0.001 | 0,59 | 0,999 |
| Phalanges+Sesamoides CAV UL – OSC LM | 51,12 | 1 | <0.001 | 0,83 | 1 |

**Table A.16** Results of test for differences in proportion of carpal/tarsal and phalanges/sesamoides across all ungulates between Uluzzian and Late Mousterian layers of southeastern Italy (i.e. those with no direct evidence of carnivore gnawing) with the relative effect size and statistical power.

| | X-squared | df | P-value | Effect size (Cohen's h) | Power |
|---|---|---|---|---|---|
| Diaphysis CAV UL – CAV LM | 780,01 | 1 | <0.001 | 0,5 | 1 |
| Diaphysis CAV UL – OSC LM | 883,87 | 1 | <0.001 | 0,58 | 1 |
| Epiphysis CAV UL – CAV LM | 83,663 | 1 | <0.001 | 0,15 | 1 |
| Epiphysis CAV UL – OSC LM | 176,26 | 1 | <0.001 | 0,25 | 1 |
| Spongy bones CAV UL – CAV LM | 910,5 | 1 | <0.001 | 0,55 | 1 |
| Spongy bones CAV UL – OSC LM | 142,45 | 1 | <0.001 | 0,23 | 1 |

**Table A.17** Results of test for differences in proportion of diaphysis, epiphysis, and spongy bones between Uluzzian and Late Mousterian layers of southeastern Italy (i.e. those with no direct evidence of carnivore gnawing) with the relative effect size and statistical power.